\RequirePackage{fix-cm}
\documentclass[epjc3]{svjour3}  
\smartqed  
\RequirePackage{graphicx}

\usepackage[margin=2.0cm]{geometry}

\usepackage{amssymb}
\usepackage[numbers]{natbib}

\usepackage{latexsym}
\usepackage{revsymb}
\usepackage{multirow}
\usepackage{color}
\usepackage[usenames,dvipsnames,svgnames,table]{xcolor}

\usepackage{graphicx}
\usepackage{epsfig}  
\usepackage{epsf}    
\usepackage{dcolumn}
\usepackage{bm}
\usepackage{dcolumn}
\usepackage{textcomp}
\usepackage{float}
\usepackage{subfig}

\usepackage{lineno}
\modulolinenumbers[5]

\usepackage[hyperfootnotes=false]{hyperref}
\usepackage[all]{hypcap}
\usepackage{pdftricks}
\usepackage{amsmath}
\usepackage{bm}
\usepackage{amssymb}
\usepackage{amsfonts}

\begin{document}

\title{Wave-packet treatment of reactor neutrino oscillation experiments and its implications on determining the neutrino mass hierarchy}

\author{Yat-Long Chan\thanksref{e1,addr1}
        \and
        M. -C. Chu\thanksref{e2,addr1} 
        \and
        Ka Ming Tsui\thanksref{e3,addr2} 
        \and
        Chan Fai Wong\thanksref{e4,addr3} 
        \and
        Jianyi Xu\thanksref{e5,addr1} 
}

\thankstext{e1}{ylchan87@gmail.com}
\thankstext{e2}{mcchu@phy.cuhk.edu.hk}
\thankstext{e3}{kmtsui@icrr.u-tokyo.ac.jp}
\thankstext{e4}{wongchf@mail.sysu.edu.cn}
\thankstext{e5}{jyxu@phy.cuhk.edu.hk}


\institute{ Department of Physics, The Chinese University of Hong Kong, N.T., Hong Kong \label{addr1}
           \and
           RCCN, ICRR, University of Tokyo, Kashiwa-no-ha Kashiwa City, Chiba, Japan \label{addr2} 
           \and
           Sun Yat-Sen University, GuangZhou, China \label{addr3}
}

\date{}

\maketitle

\begin{abstract}
We derive the neutrino flavor transition probabilities with the neutrino treated as a wave packet. The decoherence and dispersion effects from the wave-packet treatment show up as damping and phase-shifting of the plane-wave neutrino oscillation patterns. If the energy uncertainty in the initial neutrino wave packet is larger than around 0.01 of the neutrino energy, the decoherence and dispersion effects would degrade the sensitivity of reactor neutrino experiments to mass hierarchy measurement to lower than 3 $\sigma$ confidence level.

\end{abstract}



\section{Introduction}
Much information regarding neutrino mixing have
been revealed in the past few decades. However, in most oscillation data analyses, neutrinos are described as plane waves with definite energy and momentum. 
Since neutrino production and detection are spatially localized, a wave-packet description is more general and appropriate for a complete understanding of neutrino oscillations. Even if the plane-wave treatment is a good approximation for neutrino flavor transitions, the wave-packet decoherence and dispersion effects could still give rise to small corrections to oscillation parameters. 
We investigate the wave-packet treatment in detail, constrain the energy uncertainty $\sigma_\mathrm{wp}$ of reactor anti-neutrinos, and calculate corrections to the mixing parameters by the Daya Bay \cite{DayaBay2013} and KamLAND \cite{KamLand} reactor neutrino experiments.

According to our analyses, the wave-packet treatment does not produce significant modifications of the mixing parameters measured by current reactor neutrino experiments based on the plane-wave analysis. However, current experimental data allows a large possible range in the initial momentum width of the neutrino wave packet ($\sigma_\nu$). If the initial momentum/energy uncertainty of the neutrino wave packet is larger than around 0.02 of the neutrino energy, the decoherence and dispersion effects could have significant effects on future measurements of the neutrino mass hierarchy. 

In this article, we apply a wave-packet treatment to neutrino oscillations \cite{Fuji:2006mq, Blennow:2005yk, Bernardini:2006ak,Giunti:2007ry,Naumov:2013vea, Akhmedov:2009rb, Naumov:2010um} and examine its phenomenological implications on reactor neutrino experiments at medium baseline.

The article is organized as follows: In Section 2, the wave-packet treatment is briefly reviewed. In Section 3, we use the survival probability derived in Section 2 to explore the sensitivity of potential measurements of mass hierarchy with medium baseline reactor experiments. We summarize and conclude in Section 4.

\section{Wave-packet impact on current reactor neutrino experiments}
\subsection{Wave-packet treatment for neutrino oscillations}
The plane-wave description of neutrino oscillation has been developed for almost 40 years \cite{Eliezer:1975ja}. In the standard calculation of plane-wave neutrino oscillations, the probability of detecting a neutrino flavor state $|\nu_\beta \rangle$ with energy $E$, evolved from a pure flavor state $|\nu_\alpha \rangle$, at a distance $z$ from the production point is given by
\begin{align}\label{OSC_eq}
 P_{\alpha \beta} & = \delta_{\alpha \beta} - 4\sum_{k > j}\mathrm{Re}(U_{\alpha j}U_{\beta j}^\ast U_{\alpha k}^\ast U_{\beta k})\mathrm{sin}^2(\Delta m_{kj}^2 z/4E) \notag \\ 
                  & + 2\sum_{k > j}\mathrm{Im}(U_{\alpha j}U_{\beta j}^\ast U_{\alpha k}^\ast U_{\beta k})\mathrm{sin}(\Delta m_{kj}^2 z/2E),
\end{align}
where $U_{\alpha i}$ denote the elements of the PMNS matrix and $\Delta m_{kj}^2 = m_k^2-m_j^2$ are the differences of the mass eigenvalues squared \cite{Maki:1962mu}.

However, as neutrino production and detection are spatially localized, there must be finite intrinsic energy/momentum uncertainties and a neutrino should be described by a wave packet. The wave-packet character of light has been discussed in details in Ref. \cite{Jenkins-White-FundamentalsOfOptics-1981}. Based on similar arguments as in that reference, all particles are produced and detected as wave packets. In 1976, the wave-packet nature of propagating neutrinos was proposed \cite{Nussinov:1976uw}. A wave-packet description is expected to be more general and appropriate for a complete understanding of neutrino oscillations \cite{Kayser:1981ye,Kiers:1995zj,Giunti:2002xg,Dolgov:2005vj,Giunti:2007ry,Akhmedov:2009rb,Naumov:2010um}. Nevertheless, there are also arguments against the wave-packet treatment. Refs. \cite{Stodolsky:1998tc,Lipkin:2002sq} argue that a wave-packet description is unnecessary as the oscillation system is stationary. However, it has been pointed out in Refs. \cite{Beuthe:2001rc, Giunti:2003ax, Farzan:2008eg, Bilenky:2011pk, priv:Dmitry} that the authors of Refs. \cite{Stodolsky:1998tc,Lipkin:2002sq} have mixed the macroscopic stationarity with microscopic stationarity. 
The wave-packet description of neutrino oscillations is necessary at least in principle.

\begin{figure*}
\centering
 \includegraphics[scale=0.56]{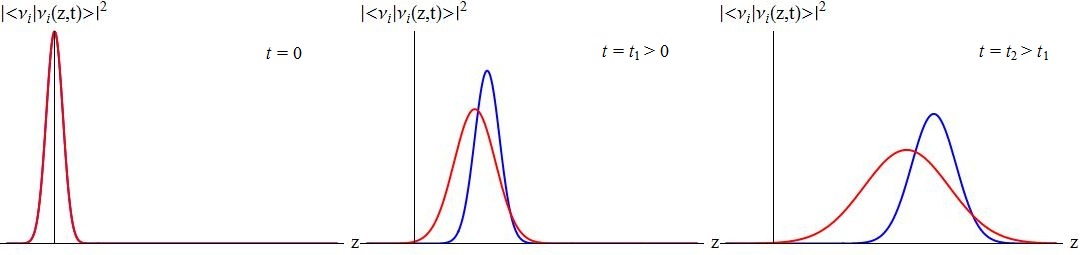}
\caption{An illustration of the decoherence and dispersion effects on neutrino flavor oscillations. The different mass eigenstates (red and blue lines) in a flavor neutrino would gradually separate in space due to speed differences, leading to decoherence and reduction of interference.  At the same time, each mass eigenstate wave packet would disperse, partially compensating for the decoherence effect. However, the overlapping fractions of the wave packets remain smaller than those of plane waves.}
\label{fig:ROCK}
\end{figure*}

Therefore, a neutrino is described by a wave packet as it propagates freely \cite{Kiers:1995zj, Akhmedov:2009rb, Naumov:2010um}:

\begin{align}\label{wave_packet}
 |\nu_i(z,t) \rangle & = \int_{- \infty}^\infty \dfrac{dp}{\sqrt{2\pi}}\dfrac{1}{\sqrt{\sqrt{\pi}\sigma_\nu}}\mathrm{exp} \left[-\dfrac{(p-p_\nu)^2}{2\sigma_\nu^2}\right] \notag \\
 & \cdot \mathrm{exp}[i(pz-E_i(p)t)] |\nu_i \rangle, \\  
 |\nu_\alpha(z,t) \rangle & = \sum_{i} U_{\alpha i}^\ast |\nu_i(z,t) \rangle, \label{flavor_state}
\end{align}
where $|\nu_i \rangle$ is an energy eigenstate with energy $E_i$, $p_\nu$ is the mean momentum, $\sigma_\nu$ is the width of the wave
packet in momentum space\footnote{Here, $\sigma_\nu$ is the effective uncertainty, with $1/\sigma_\nu^2 = 1/\sigma_\mathrm{prod}^2 + 1/\sigma_\mathrm{det}^2$, which has included both the production and detection neutrino energy uncertainties \cite{Giunti:2002xg, Giunti:2003ax, Beuthe:2001rc}. Moreover, we would like to point out that $\sigma_\mathrm{det}$ represents the energy uncertainty of detection at the microscopic level, ie., that of the inverse-beta decay reaction. This is different from the detector energy resolution, which is determined by macroscopic parameters such as the performance of PMTs and geometry of the anti-neutrino detector, etc. In principle, the detector resolution is irrelevant for the size of the neutrino wave packets.}, assumed to be independent of the neutrino energy here, and $|\nu_\alpha \rangle$ is a neutrino flavor state. Fig. \ref{fig:ROCK} pictorially describes the wave-packet effects on the propagations of neutrino mass eigenstates.

In order to calculate the integral in Eq. (\ref{wave_packet}), we expand the energy $E_i(p)$ around the mean momentum $p_\nu$ up to second order
\begin{equation}\label{E_expand}
 E_i(p) \approx E_i(p_\nu) + v_i(p_\nu)(p-p_\nu) + \dfrac{m_i^2}{2(E_i(p_\nu))^3}(p-p_\nu)^2,
\end{equation}
where $v_i(p_\nu) = \left. \dfrac{dE_i}{dp}\right|_{p = p_\nu} = p_\nu/E_i (p_\nu)$, is the group velocity of wave packet.  We use Eq. (\ref{wave_packet}), (\ref{flavor_state}) and (\ref{E_expand}) to calculate the neutrino flavor transition probabilities at baseline $L$:  
\begin{align}
 P_{\nu_\alpha \rightarrow \nu_\beta}(L) & \approx \sum_{ij} \left\{ U_{\alpha i}^\ast U_{\beta i}U_{\alpha j}U_{\beta j}^\ast \mathrm{exp}\left[-i\dfrac{2\pi L}{L^\mathrm{osc}_{ij}}\right] \right\} \notag \\
 & \left\{ \left(\dfrac{1}{1+y_{ij}^2}\right)^{\frac{1}{4}}\mathrm{exp}(-\lambda_{ij})\mathrm{exp}\left(\frac{-i}{2} \mbox{tan}^{-1} (y_{ij})\right) \mathrm{exp}(i\lambda_{ij}y_{ij}) \right\}, \label{Prob_inf1}  \\
                                         \mbox{where }& \lambda_{ij} \equiv \dfrac{x_{ij}^2}{1+y_{ij}^2},  \quad \qquad y_{ij} \equiv \dfrac{L}{L^\mathrm{dis}_{ij}},  \quad \qquad x_{ij} \equiv \dfrac{L}{L^\mathrm{coh}_{ij}}, \notag \\
                                        & L^\mathrm{coh}_{ij} \equiv \dfrac{L^\mathrm{osc}_{ij}}{\pi \sigma_\mathrm{wp}}, \qquad \quad L^\mathrm{dis}_{ij} \equiv \dfrac{L^\mathrm{osc}_{ij}}{2\pi\sigma_\mathrm{wp}^2}, \qquad L^\mathrm{osc}_{ij} \equiv \dfrac{4\pi E}{\Delta m^2_{ij}}, \notag \\
                                        & \sigma_\mathrm{wp} = \dfrac{\sigma_\nu}{E_i(p_\nu)} \approx \dfrac{\sigma_\nu}{E(p_\nu)} \notag.
\end{align}
Detailed derivation of Eq. (\ref{Prob_inf1}) is shown in Appendix A. In Eq. (\ref{Prob_inf1}), the terms in the first bracket correspond to the standard plane-wave oscillation probabilities, and those in the second bracket represent the modifications from the wave-packet treatment. The exp$(-\lambda_{ij})$ term corresponds to the decoherence effect due to the fact that different mass states propagate at different speeds $v_i(p_\nu)$ and they gradually separate and stop to interfere with each other, resulting in a damping of oscillations. The terms depending on $y_{ij}$ come from the quadratic correction in Eq. (\ref{E_expand}); they describe the dispersion effects and are dependent on the dispersion length(s)\footnote{The ``dispersion length'' in this report represents the distance where the dispersion effect becomes important in the neutrino oscillation. A different definition of dispersion length and more detailed discussion of dispersion can be found in \cite{Beuthe:2001rc, Naumov:2013vea, Dolgov:2005vj}.} $L^\mathrm{dis}_{ij}$. Furthermore, 
$y_{ij}$ are proportional to $\sigma_\mathrm{wp}^2$, while $x_{ij} \propto \sigma_\mathrm{wp}$ only. Therefore, if $\sigma_\mathrm{wp} \ll 1$, the dispersion effect is expected to be more suppressed and negligible
. Dispersion has two effects on the oscillations. On the one hand the spreading of the wave packet compensates for the spatial separation of the mass states, hence restoring parts of their interferences. On the other hand, dispersion reduces the overlapping fraction of the wave packets, and thus the interference or oscillation effects cannot be fully restored. 
Moreover, it also modifies the flavor oscillation phases:
\begin{equation}\label{def_phi}
\phi_{ij} \equiv \dfrac{2\pi L}{L^\mathrm{osc}_{ij}} + \left(\frac{1}{2} \mbox{tan}^{-1} (y_{ij}) - \lambda_{ij}y_{ij} \right),
\end{equation}
with deviations from the standard plane-wave oscillation phase written in the parentheses. If $y_{ij}$ = 0, then $\phi_{ij}$ just reduce to the standard plane-wave oscillation phases.

Additional discussions about the phenomenological consequences of the wave-packet treatment and details of the derivation of Eq. (\ref{Prob_inf1}) can be found in \cite{OurPaper, Bernardini:2006ak, Naumov:2013vea, Beuthe:2001rc, Akhmedov:2009rb}. 

In this paper, we focus on the analyses of reactor neturino experiments. According to our wave-packet treatment, in reactor neutrino experiments, the anti-electron neutrino survival probability is

\begin{align}\label{Pee_app}
 P_{\bar{e}\bar{e}} = 1 - & \frac{1}{2}\mathrm{cos}^4(\theta_{13})\mathrm{sin}^2(2\theta_{12})[1 - (\dfrac{1}{1+y_{21}^2})^{\frac{1}{4}}\mathrm{exp}(-\lambda_{21})\mathrm{cos}(\phi_{21})] - \notag \\
          & \frac{1}{2}\mathrm{sin}^2(2\theta_{13})\mathrm{cos}^2(\theta_{12})[1 - (\dfrac{1}{1+y_{31}^2})^{\frac{1}{4}}\mathrm{exp}(-\lambda_{31})\mathrm{cos}(\phi_{31})] - \notag \\
          & \frac{1}{2}\mathrm{sin}^2(2\theta_{13})\mathrm{sin}^2(\theta_{12})[1 - (\dfrac{1}{1+y_{32}^2})^{\frac{1}{4}}\mathrm{exp}(-\lambda_{32})\mathrm{cos}(\phi_{32})].
\end{align}

\subsection{The constraints from current reactor experiments}
\begin{figure}[!htbp]
\centering
 \includegraphics[scale=0.43]{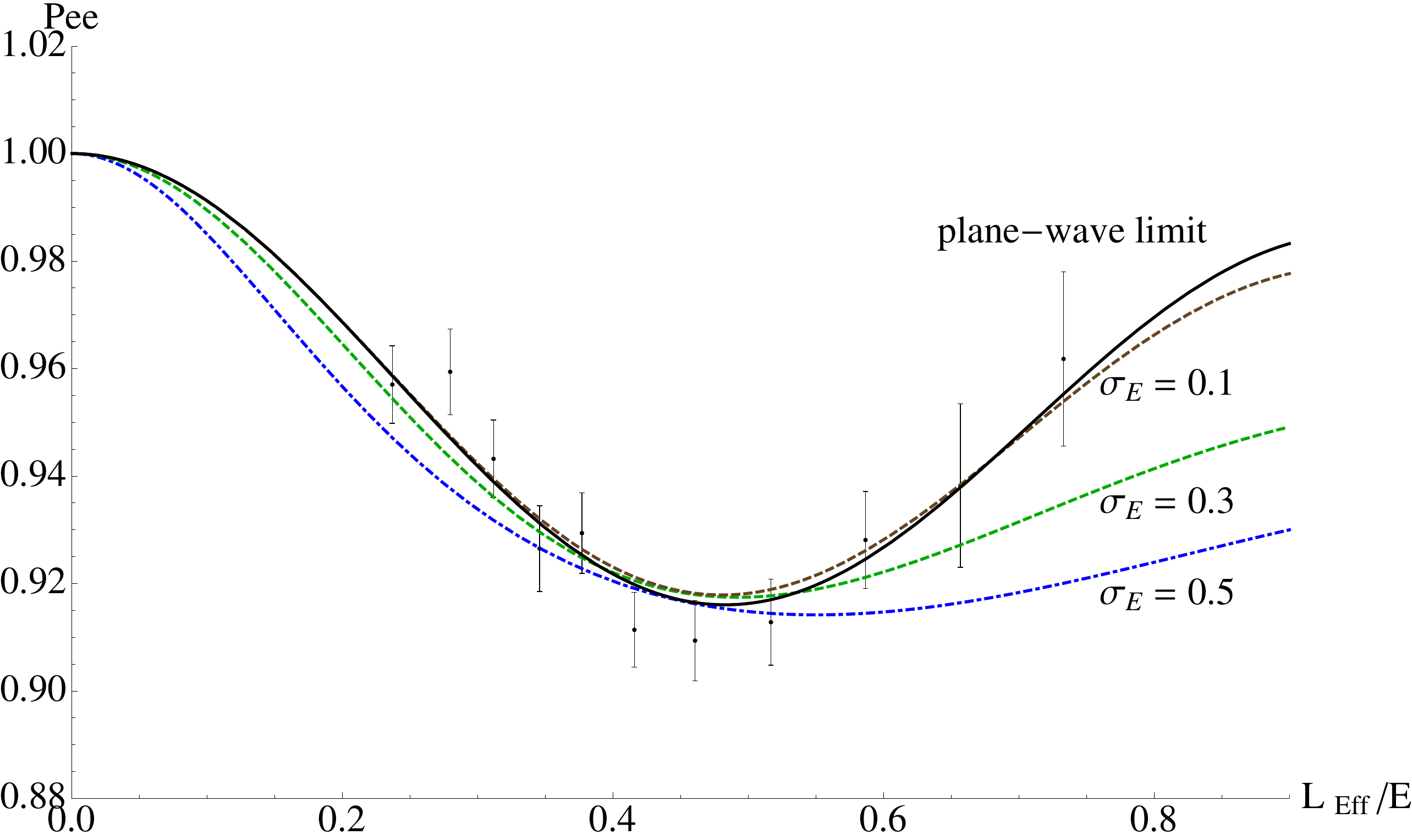}
\caption{$P_{\bar{e}\bar{e}}$ as a function of $L_\mathrm{effective}/E$ for Daya Bay \textbf{EH3}, compared with data \cite{An:2015rpe}. The black solid curve is the standard plane-wave result (sin$^2 2\theta_{13}$ = 0.084, $\Delta m^2_{32}$ = $2.43\times 10^{-3}$ eV$^2$, $\sigma_\mathrm{wp}$ is negligible), while the brown, green and blue curves are wave-packet results with $\sigma_\mathrm{wp}$ = 0.1 (and the corresponding best-fit values of sin$^2 2\theta_{13}$ = 0.084, $\Delta m^2_{32}$ = $2.4\times 10^{-3}$ eV$^2$), 0.3 (and the corresponding best-fit values of sin$^2 2\theta_{13}$ = 0.96, $\Delta m^2_{32}$ = $2.22\times 10^{-3}$ eV$^2$) and 0.5 (and the corresponding best-fit values of sin$^2 2\theta_{13}$ = 0.108, $\Delta m^2_{32}$ = $2.10\times 10^{-3}$ eV$^2$) respectively. $L_\mathrm{effective}$ was obtained by equating the actual flux to an effective antineutrino flux using a single baseline.}
\label{EH3}
\end{figure}

\begin{figure}[!htbp]
\centering
 \includegraphics[scale=0.35]{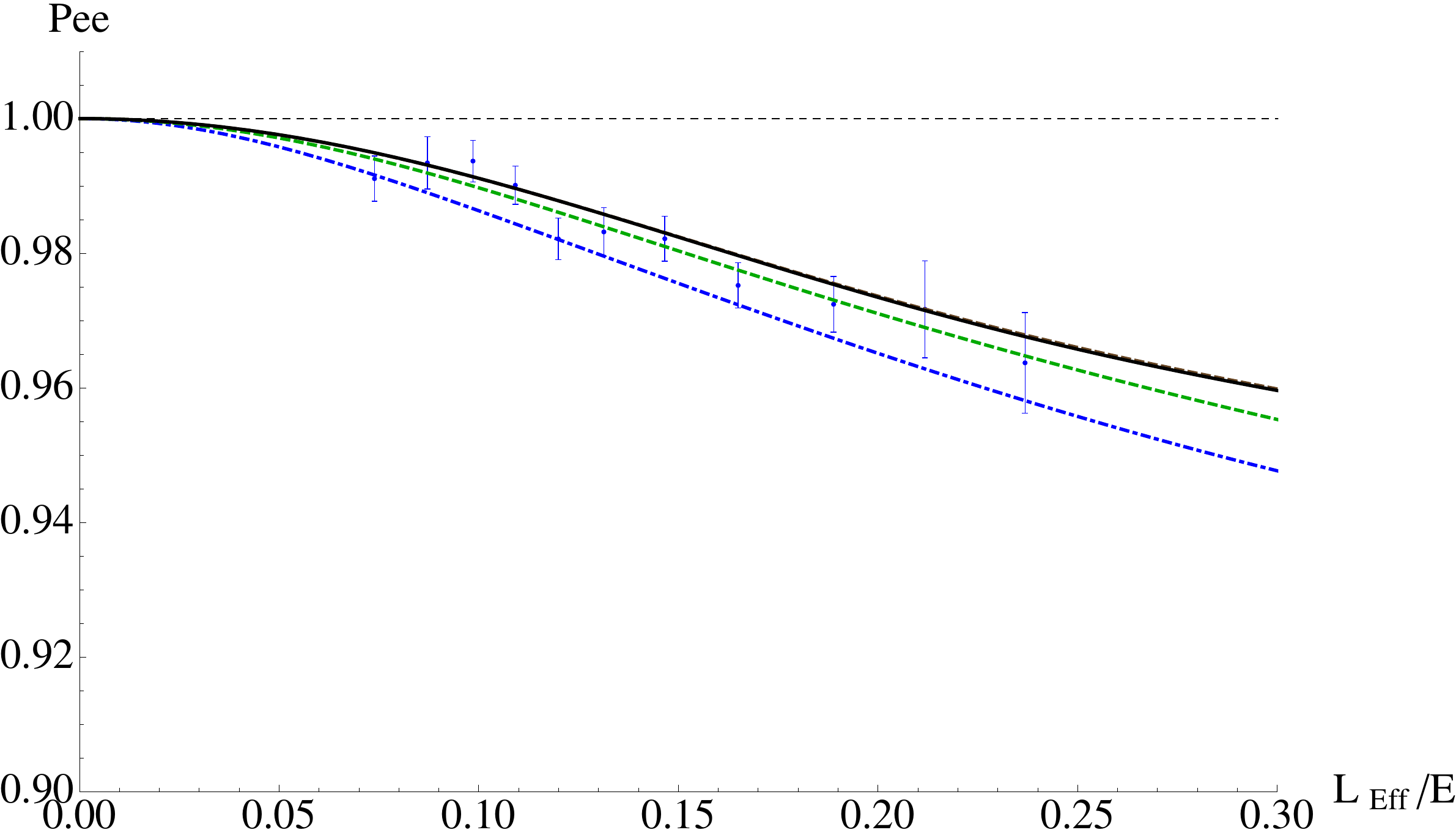} \\
 \includegraphics[scale=0.35]{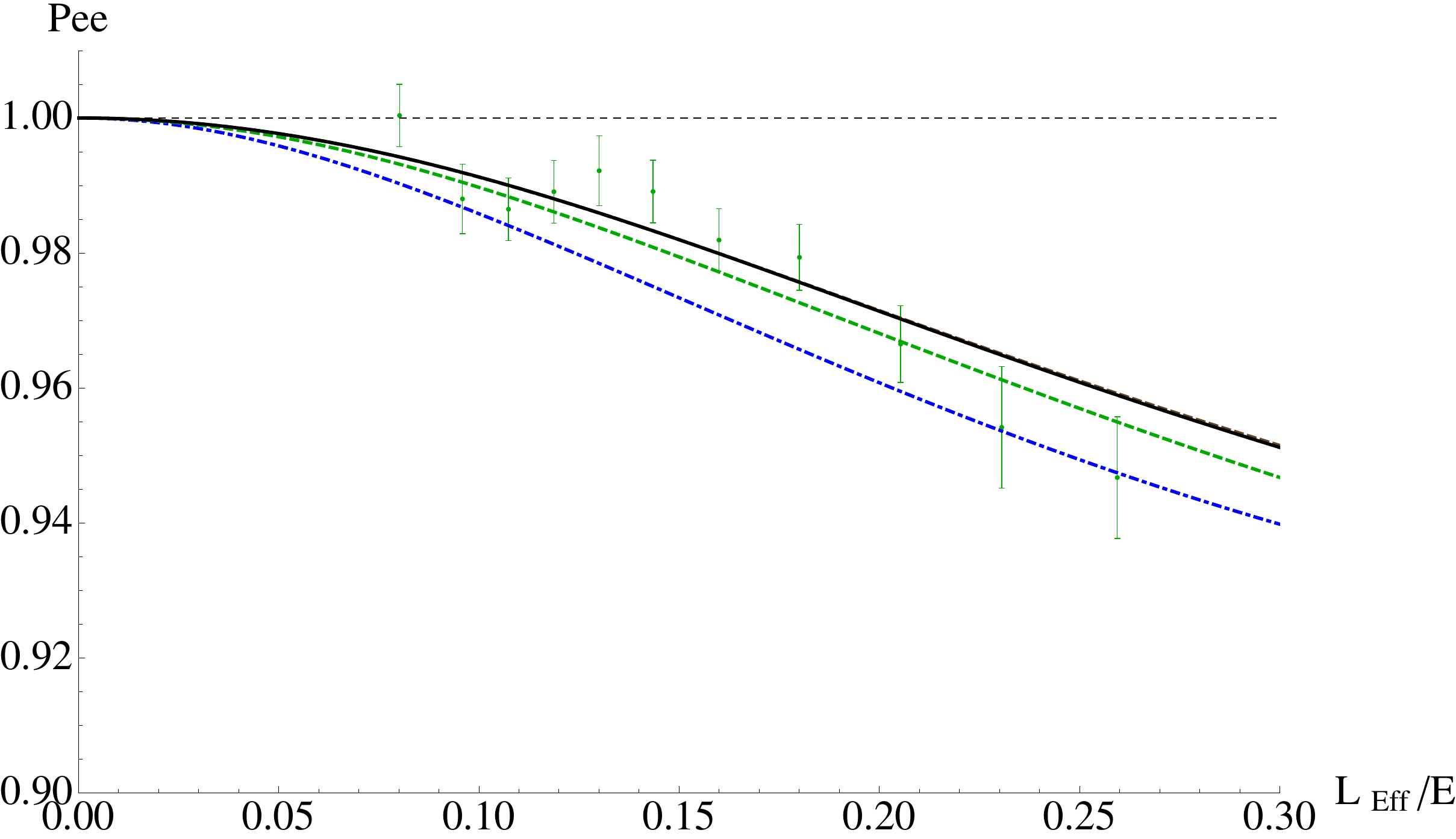}
\caption{Same as Fig. \ref{EH3}, but for Daya Bay \textbf{EH1} (upper) and \textbf{EH2} (lower). }
\label{EH1EH2}
\end{figure}

\begin{figure}[!htbp]
\centering
 \includegraphics[scale=0.35]{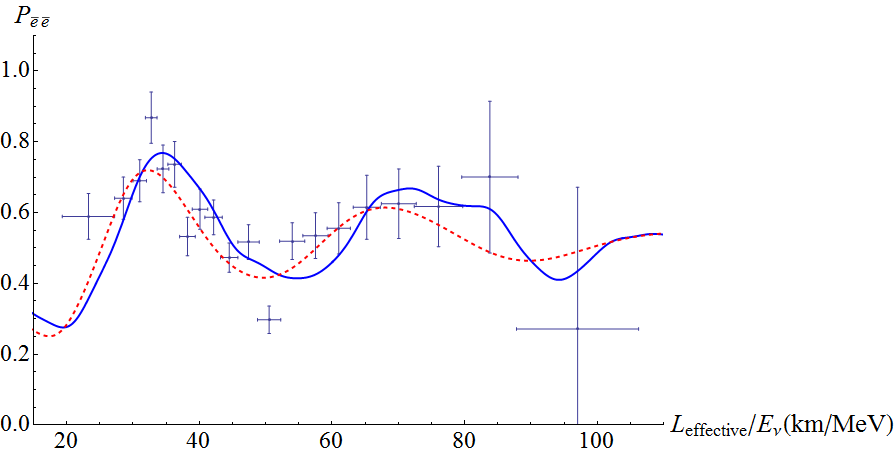}
\caption{$P_{\bar{e}\bar{e}}$ as a function of $L_\mathrm{effective}/E$ in the KamLAND experiment. The data points are the ratios of the geo-neutrino and background subtracted $\bar{\nu_e}$ spectrum to the expected no-oscillation spectrum \cite{KamLand}. $L_\mathrm{effective}$ = 180 km is the flux-weighted average reactor baseline. The blue (solid) curve is the best-fit plane-wave result, with sin$^2 2\theta_{12}$ = 0.857 and $\Delta m^2_{21}$ = $7.53\times 10^{-5}$ eV$^2$; while the red (dashed) curve is the best-fit wave-packet result, with $\sigma_\mathrm{wp}$ = 0.12, sin$^2 2\theta_{12}$ = 0.902 and $\Delta m^2_{21}$ = $8.06\times 10^{-5}$ eV$^2$. The error bars show the statistical uncertainties only.}
\label{KamLAND_curves}
\end{figure}

To date, the value of the parameter $\sigma_\mathrm{wp}$ has not yet been determined. If $\sigma_\mathrm{wp}$ is not negligible, the wave-packet effects could be significant and have important implications on current and future neutrino oscillation experiments. In this article, we constrain $\sigma_\mathrm{wp}$ by analyzing the published Daya Bay and KamLAND data shown in references \cite{An:2015rpe} and \cite{KamLand}, considering statistical errors only. Figs. \ref{EH3} and \ref{EH1EH2} show the data points from Daya Bay experiment \cite{An:2015rpe}, along with the oscillation curves corresponding to different values of $\sigma_\mathrm{wp}$. Fig. \ref{KamLAND_curves} shows the data points from KamLAND \cite{KamLand} and the oscillation curves of plane-wave and wave-packet treatments. In Figs. \ref{EH3} to \ref{KamLAND_curves}, $L_\mathrm{effective}$ are the flux-weighted average reactor baselines\footnote{In Daya Bay, the effective baselines are calculated for all three experimental halls.}, and the error bars just show the statistical uncertainties. 

We use Eq. (\ref{Pee_app}) to fit the data points in Figs. \ref{EH3} and \ref{EH1EH2} to get the constraint of $\sigma_\mathrm{wp}$ from the Daya Bay Experiment. In these figures, the black (solid) curves correspond to $\sigma_\mathrm{wp}$ = 0, sin$^2 2\theta_{13}$ = 0.084 and $\Delta m^2_{32}$ = $2.43\times 10^{-3}$ eV$^2$, same as those of the standard plane-wave treatment. The brown (dashed) curve represents the wave-packet result with $\sigma_\mathrm{wp}$ = 0.1, sin$^2 2\theta_{13}$ = 0.084 and $\Delta m^2_{32}$ = $2.4\times 10^{-3}$ eV$^2$ (the best-fit mixing parameters when $\sigma_\mathrm{wp}$ = 0.1). The green (dashed) curve shows the oscillation pattern with $\sigma_\mathrm{wp}$ = 0.3, sin$^2 2\theta_{13}$ = 0.096 and $\Delta m^2_{32}$ = $2.22\times 10^{-3}$ eV$^2$ (the best-fit mixing parameters when $\sigma_\mathrm{wp}$ = 0.3).  The blue (dot-dashed) curve corresponds to $\sigma_\mathrm{wp}$ = 0.5, sin$^2 2\theta_{13}$ = 0.108 and $\Delta m^2_{32}$ = $2.10\times 10^{-3}$ eV$^2$ (the best-fit mixing parameters when $\sigma_\mathrm{wp}$ = 0.5). The data points in Figs. \ref{EH3} and \ref{EH1EH2} show clearly the existence of neutrino oscillation, and the data agree with the $P_{\bar{e}\bar{e}}$ derived by the plane-wave approach (Eq. (\ref{OSC_eq})) for a certain set of mixing parameters. This is not surprising since the baselines of the Daya Bay Experiment are short compared to the coherence length
, so that the oscillations are not washed out yet.

Moreover, the decoherence and dispersion effects are dependent on the baseline $L$, and thus the wave-packet impact in the far hall of Daya Bay is expected to be more significant than in the near halls. In the plots in Fig. \ref{EH1EH2}, the black (solid) and brown (dashed) curves overlap almost completely, while the differences between the black, green and blue curves are also small compared to the error bars. However, for the far hall EH3, the wave-packet impact becomes significant for $\sigma_\mathrm{wp}$ $\geq$ 0.3. It is because in Eq. (\ref{Pee_app}), the term $\lambda_{ij}$ depends on the baseline $L$. Since the effective baseline of EH3 is longer, the damping of oscillation (decoherence effect) in EH3 is more significant. 

\begin{figure}[!htbp]
\centering
 \includegraphics[scale=0.5]{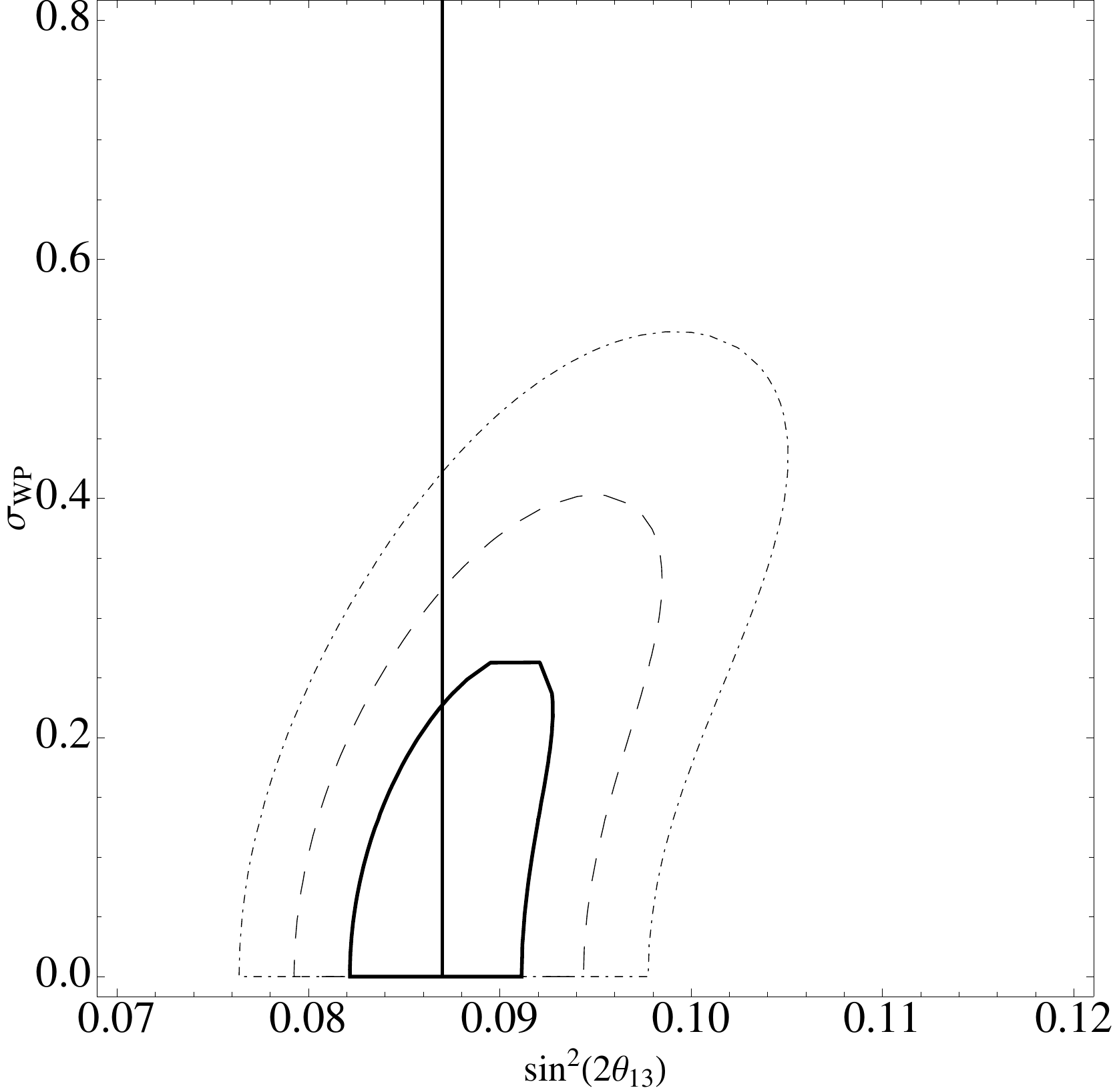}
\caption{The 1 $\sigma$ (solid), 2 $\sigma$ (dashed) and 3 $\sigma$ (dotdashed) constraints on ``$\sigma_\mathrm{wp}$ vs sin$^2 2\theta_{13}$'' from fitting the published Daya Bay data \cite{An:2015rpe}. Only statistical errors are considered. We have marginalized over $\Delta m^2_{31}$ in this plot. The vertical black line represents the best-fit sin$^2 2\theta_{13}$  based on the plane-wave analysis.}
\label{DB_t13_sigmaE}
\end{figure}

The result of our data analysis is shown in Fig. \ref{DB_t13_sigmaE}. Here, we have just considered the statistical errors. The constraints on the parameters could become worse with the systematic errors taken into account. Fig. \ref{DB_t13_sigmaE} shows that the wave-packet impact is not significant in Daya Bay experiment and it hardly affects the measurement of $\theta_{13}$. The vertical black line in the figure represents the best-fit value of sin$^2 2\theta_{13}$ in the plane-wave model. A larger value of $\sigma_\mathrm{wp}$, or larger decoherence effect, implies that the true value of $\theta_{13}$ should be larger\footnote{If $\sigma_\mathrm{wp}$ is non-negligible but we still see oscillation effect from the data, it means that the true value of the mixing angle is actually larger than we expected in plane-wave assumption.}. From our analysis, there is no strong evidence to suggest non-zero $\sigma_\mathrm{wp}$. Moreover, our result suggests that the modification of $\theta_{13}$ is not significant even when the wave-packet framework is considered. The best-fit value of sin$^2 2\theta_{13}$ from plane-wave analysis (vertical black line) is not ruled out even with 1 $\sigma$ C.L. We believe that it is because the effective baselines of the Daya Bay Experiment are short compared to the coherence length.    

\begin{figure}[!htbp]
\centering
 \includegraphics[scale=0.53]{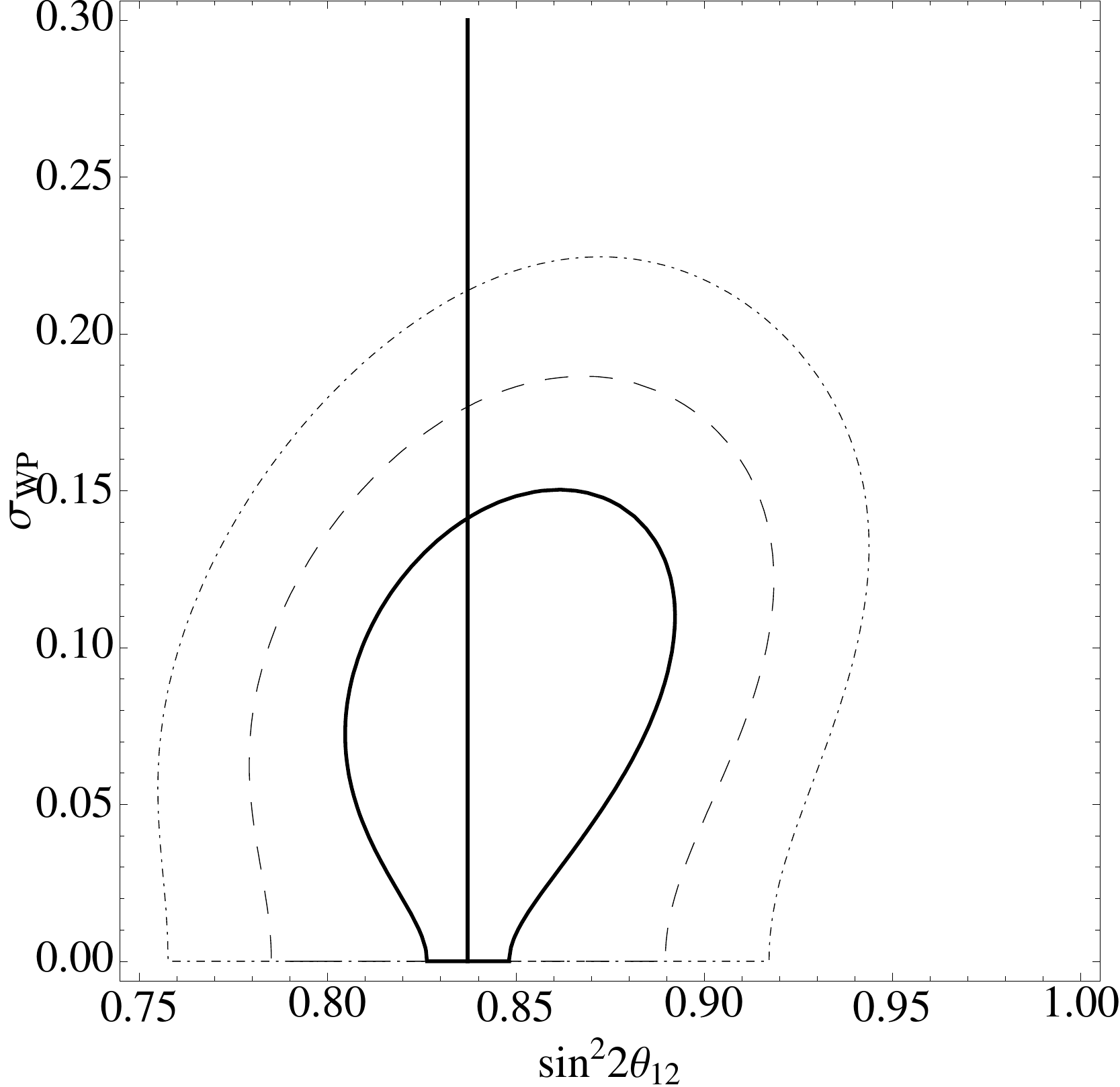}
\caption{The 1 $\sigma$ (solid), 2 $\sigma$ (dashed) and 3 $\sigma$ (dotdashed) allowed regions on ``$\sigma_\mathrm{wp}$ vs sin$^2 2\theta_{12}$'' from fitting the published KamLAND data \cite{KamLand}. Only statistical errors are considered. $\Delta m^2_{21}$ has been marginalized in this plot. The vertical black line represents the best-fit sin$^2 2\theta_{12}$ from the plane-wave analysis.}
\label{KamLAND_t12_sigmaE}
\end{figure}

We perform the similar analysis with KamLAND data from Fig. \ref{KamLAND_curves}. The result of our data analysis is shown in Fig. \ref{KamLAND_t12_sigmaE}. Again, the systematic errors are not considered.

Eq. (\ref{Prob_inf1}) shows that $L^\mathrm{coh}/L^\mathrm{dis} = 2\sigma_\mathrm{wp}$, which implies that the dispersion length is much longer than coherence length if $\sigma_\mathrm{wp}$ is negligible. This means that the decoherence effect from separation of wave packets is expected to be more significant than the dispersion effect. However, the dispersion effect is also important because it partly restores the oscillation, especially in the case of large $\sigma_\mathrm{wp}$. If the dispersion effect is not considered, the modifications of the true values of mixing angles in Figs. \ref{DB_t13_sigmaE} and \ref{KamLAND_t12_sigmaE} would be more significant. The bounds in Figs. \ref{DB_t13_sigmaE} and \ref{KamLAND_t12_sigmaE} come from the combination of decoherence and dispersion effects, but the contribution from decoherence is expected to be dominant.

Figs. \ref{DB_t13_sigmaE} and \ref{KamLAND_t12_sigmaE} show that wave-packet effects are not significant in the current reactor neutrino experiments. The 1 $\sigma$ upper bound of the energy uncertainty $\sigma_\mathrm{wp}$ $\sim$ $O(0.1)$, which is larger than some previous theoretical estimations ($\sigma_\mathrm{wp} \sim O(10^{-7})$) \cite{SigmaxRange, Giunti:2007ry, Akhmedov:2009rb, Giunti:2006fr}. Although our analyses have not considered systematic errors, our study on the current reactor experiments suggest that $\sigma_\mathrm{wp}$ can be around $O(0.1)$ for reactor experiments. We emphasize that $\sigma_{wp}$ in this article is an effective parameter which include both the production and detection neutrino energy uncertainties. 
The estimation of the value of this parameter or the size of neutrino wave packet has not come to a strong conclusion yet. Conventionally, it has been argued that $\sigma _{wp}$ should be much smaller than 1 \cite{Giunti:2006fr}. However, there are still no experimental support for such an assumption. In this paper we do not calculate or suggest the theoretical value of $\sigma _{wp}$. 
We point out that the wave-packet impact is not significant for current reactor neutrino experiments. Nevertheless, the 1 $\sigma$ C.L. allowed range of $\sigma_\mathrm{wp}$ is $O(0.1)$, within which the potential wave-packet impact could lead to significant effects and additional challenges in future neutrino oscillation experiments.

\section{Measuring neutrino mass hierarchy in reactor neutrino experiments}
\subsection{Measurement of neutrino mass hierarchy}
The signs for $\Delta m^2_{31}$ and $\Delta m^2_{32}$ have 
not yet been determined. \textbf{Normal Hierarchy} (NH) corresponds to positive $\Delta m^2_{31}$ and $\Delta m^2_{32}$ with $\nu_1$ as the lightest mass state. \textbf{Inverted Hierarchy} (IH) corresponds to negative $\Delta m^2_{31}$ and $\Delta m^2_{32}$ with $\nu_3$ as the lightest mass state \cite{TheoryOfNu}. 

As indicated by the recent data obtained by the Daya Bay Experiment, $\mathrm{sin}^2 2\theta_{13} = 0.084 \pm 0.005$ \cite{An:2015rpe}. With such a (relatively) large value of $\theta_{13}$, it is possible to determine the neutrino mass hierarchy (MH) in reactor neutrino experiments at medium baseline \cite{Petcov, Yifang}.

For a detector at baseline $L$, the observed anti-electron neutrino flux of visible energy $E_\mathrm{vis}$ is given by
\begin{align}\label{nu_flux}
 f (E_\mathrm{vis}) & = \dfrac{1}{4\pi L^2}\int dE \phi(E) \sigma (E) P_{\bar{e}\bar{e}}(L,E) \notag \\
                                   & \cdot R(E-0.8\mathrm{MeV} - E_\mathrm{vis}, \delta E).
\end{align}
$\phi(E)$ is the reactor neutrino energy spectrum, and $\sigma(E)$ is the inverse beta decay cross section. $R$ is the detector response function with energy resolution $\delta E$, which will be discussed in more detail in the next subsection.

\begin{figure}[!htbp]
\centering
{ \includegraphics[scale=0.43]{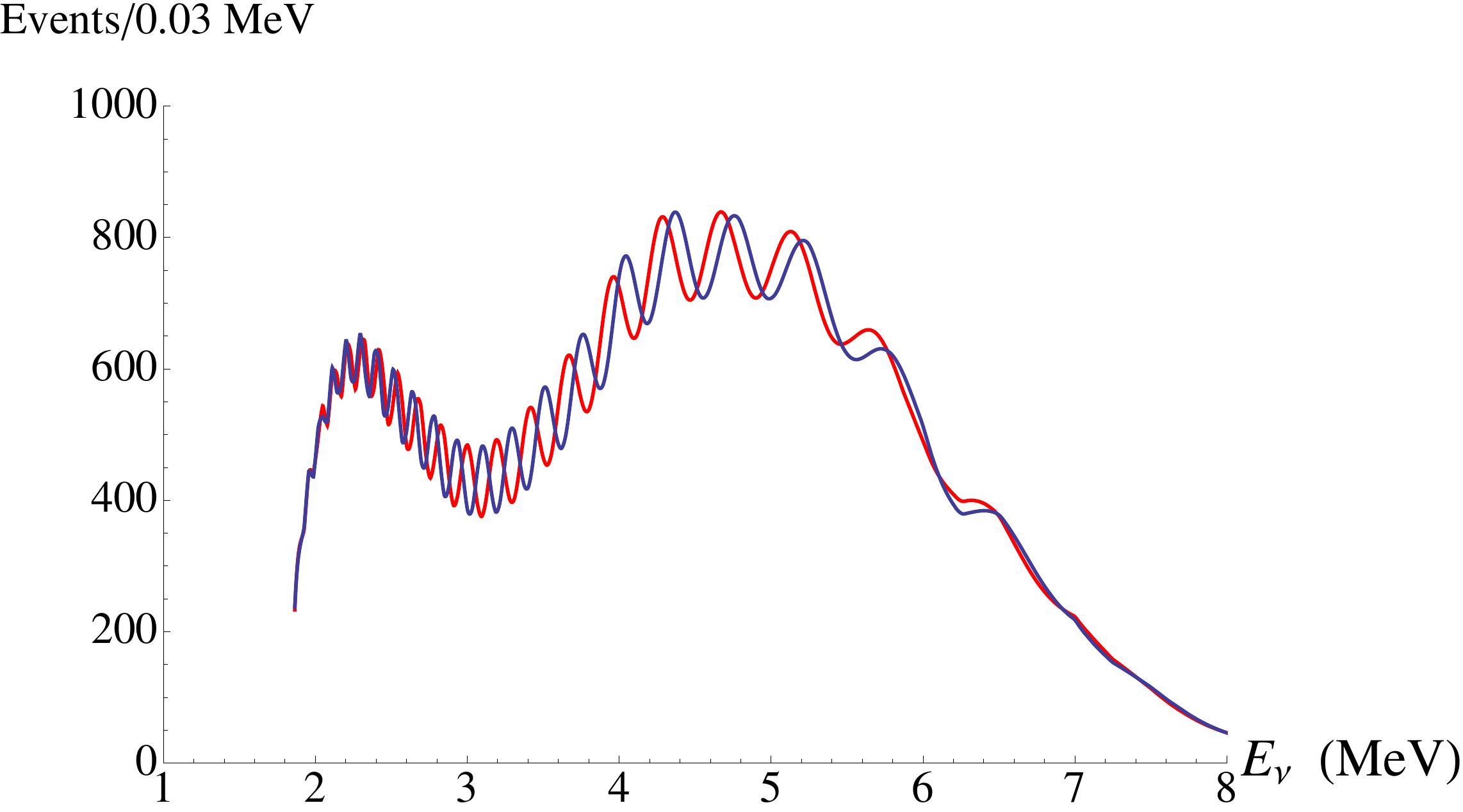} \\
 \includegraphics[scale=0.4]{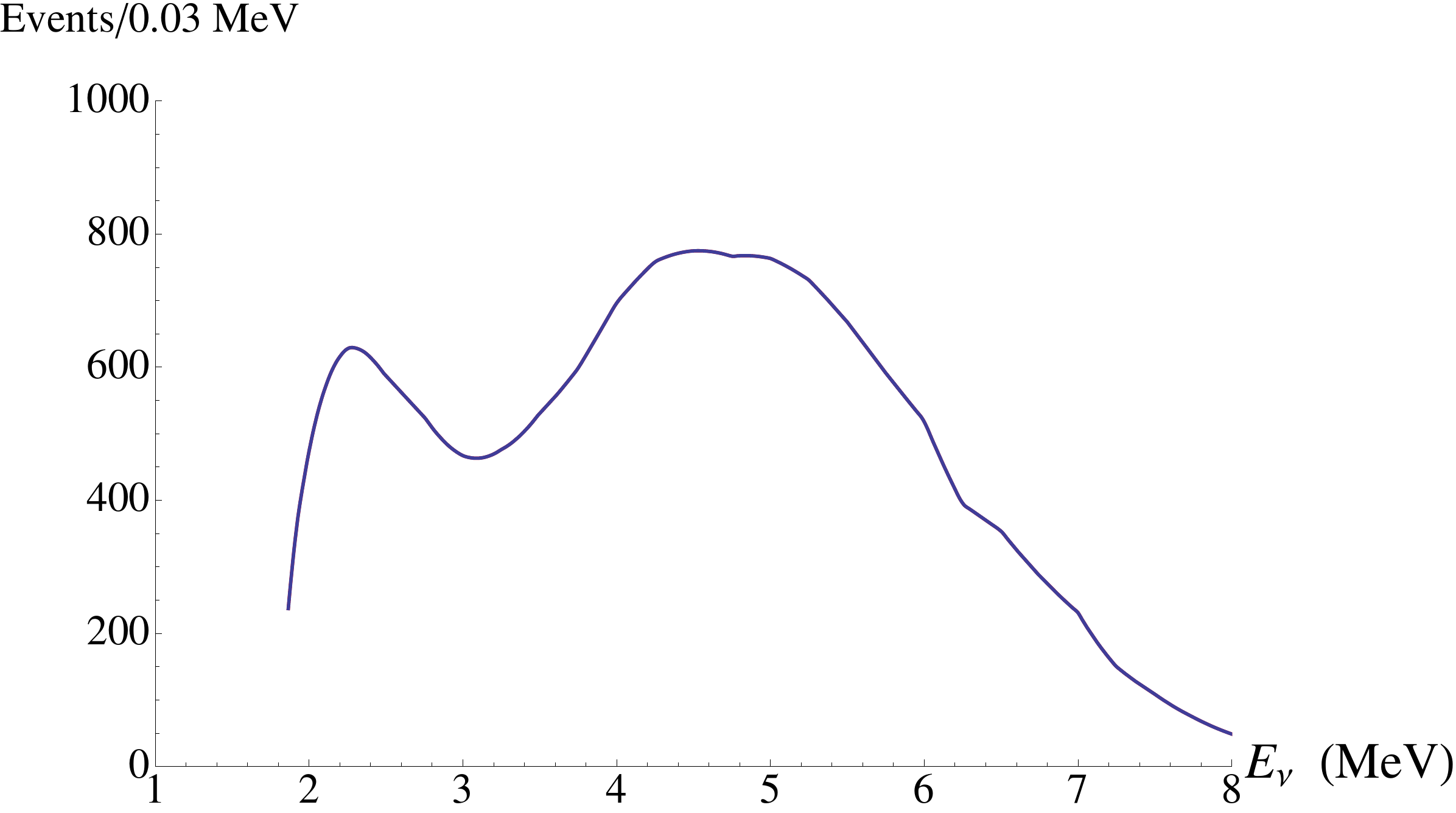}
} 
\caption{Upper panel: Expected neutrino visible energy spectrum for plane wave neutrino oscillation at 53 km for NH (solid) and IH (dashed). \\
Lower panel: Same as in upper panel, but with $\sigma_\mathrm{wp}$ = 0.1. The two curves for NH and IH completely overlap due to the wave-packet impacts.}
\label{58km_flux}
\end{figure}

It is known that at a baseline of around 50 km, which corresponds to the first minimum of $\theta_{12}$ oscillation for reactor neutrinos, the sensitivity for measuring MH is maximal. The upper panel in Fig. \ref{58km_flux} shows $f (E_\mathrm{vis})$ at $L$ = 53 km for NH and IH, with standard oscillation parameters in the plane-wave treatment ($\sigma_\mathrm{wp} = 0$ in Eq. (\ref{Pee_app})). 

\subsection{Impact of wave-packet treatment}

Our wave-packet treatment shows that the amplitudes of neutrino oscillations will be reduced. 
In particular, damping of the $\theta_{13}$ oscillations will be significant for an intermediate baseline reactor neutrino experiment if $\sigma_\mathrm{wp}$ is $O (0.1)$. The lower panel in Fig. \ref{58km_flux} shows the neutrino visible energy spectrum at a baseline of 53 km, with standard neutrino mixing parameters and $\sigma_\mathrm{wp}$ = 0.1.
If $\sigma_\mathrm{wp}$ is large, the neutrino spectra for NH and IH are indistinguishable from each other.

We modify the GLoBES software \cite{GLoBES_04, GLoBES_07} to perform numerical simulations of a 53 km baseline reactor neutrino experiment, using a similar setup as in \cite{Qian:2012xh} and \cite{Yifang}: a 20 kton detector with 3\% energy resolution, reactors with a total thermal power of 40 GW and a nominal running time of six years. In the absence of oscillations, the total number of events is about 10$^6$. 
As this paper focuses on the wave-packet impact, the systematic errors were not taken into account in the following simulations\footnote{We have also performed simulations with the following systematic errors \cite{Yifang}: 2\% correlated reactor uncertainty, 0.8\% uncorrelated reactor uncertainty, 1\% spectral uncertainty, and 1\% detector-related uncertainty. The results are similar to what we present here.}. We took the oscillation parameter values from global analysis \cite{Global_analysis} as $\Delta m^2_{21}$ = 7.54 $\times$ 10$^{-5}$ eV$^2$, ($\Delta m^2_{31}$ 
+ $\Delta m^2_{32}$)/2 = 2.43 $\times$ 10$^{-3}$ eV$^2$, sin$^2\theta_{12}$ = 0.307 and sin$^2\theta_{13}$ = 0.0241.

To distinguish between NH and IH, we quantify the sensitivity of the MH measurement by employing the least-squares method, based on a $\chi^2$ function: 

\begin{equation}\label{chi2}
 \chi^2 = \sum^{N_\mathrm{bins}}_{i=1}\dfrac{(N_i^M - N_i^T)^2}{N_i^M},
\end{equation}
where $N_i^M$ is the measured neutrino events in the $i$th energy bin, and $N_i^T$ is the predicted number of neutrino events with oscillations taken into account (without considering the systematic errors)\footnote{Without loss of generality, in our simulations, NH is assumed to be the true mass hierarchy. The result is identical if we assume IH to be true.}. 
The number of bins used $N_\mathrm{bins}$ is 164, equally spaced between 1.8 and 10 MeV.

\begin{figure}[!htbp]
\centering{
 \includegraphics[scale=0.47]{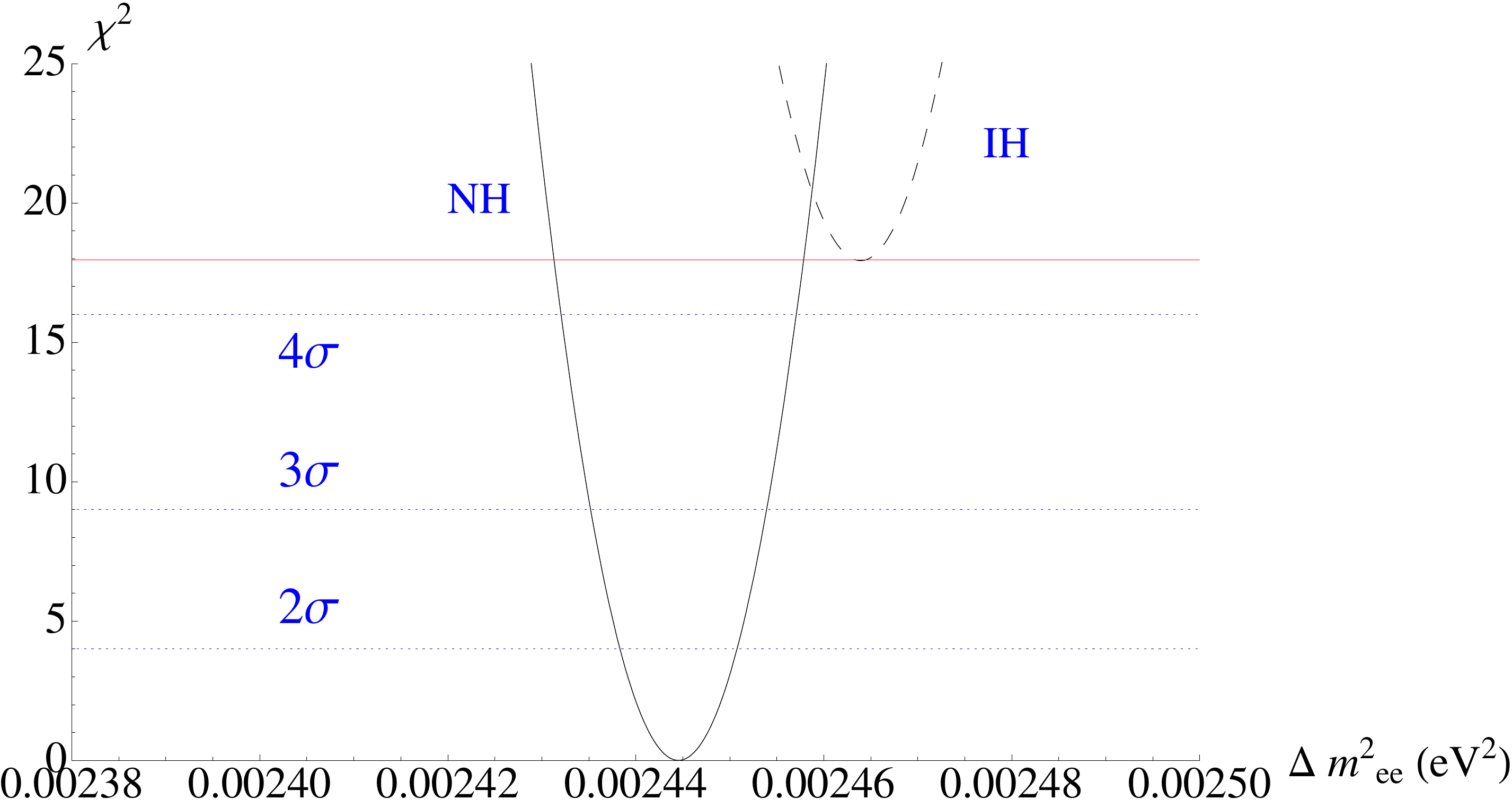} \\[0.5cm]
 \includegraphics[scale=0.47]{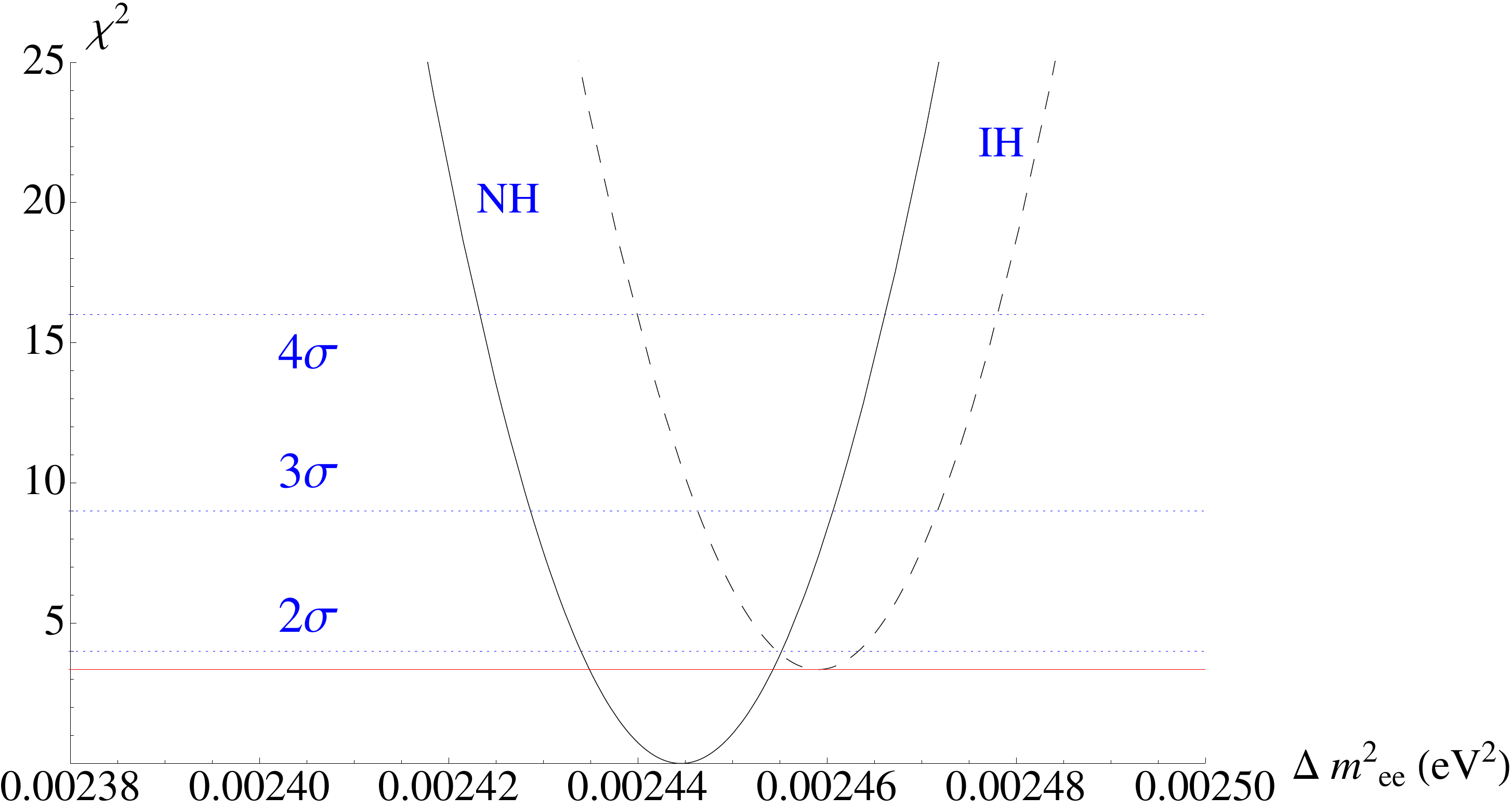} 
} 
\caption{The sensitivities of measuring $\Delta m^2_{ee}$ with and without wave-packet effects. NH is assumed to be the true hierarchy in the simulation. The solid (dashed) curve corresponds to the fitting of NH (IH, false hierarchy). $\sigma_\mathrm{wp}$ is assumed to be 0 in the upper panel and 0.02 in the lower panel.}
\label{Dchi_SigmaE}
\end{figure}

We fit the hypothetical data set with $\Delta m^2_{ee}$ as the free variable, defined as
\begin{equation}\label{dmee}
 \Delta m^2_{ee} = \mathrm{cos}^2\theta_{12} \Delta m^2_{31} + \mathrm{sin}^2\theta_{12} \Delta m^2_{32}.
\end{equation}
The capability to resolve the mass hierarchy is then given by the difference between the minimum $\chi^2$ value for IH and NH:
\begin{equation}\label{deltachi2}
 \Delta \chi^2_\mathrm{MH} = \mathrm{min}(\chi^2_\mathrm{IH}) - \mathrm{min}(\chi^2_\mathrm{NH}).
\end{equation}
$\Delta \chi^2_\mathrm{MH}$ is used to explore the wave-packet effects as well as the impact of statistics and systematics in measuring the MH.\footnote{Note that $\mathrm{min}(\chi^2_\mathrm{NH})$ and $\mathrm{min}(\chi^2_\mathrm{IH})$ can be located at different values of $|\Delta m^2_{ee}|$.} If $\sigma_\mathrm{wp} = 0$, which corresponds to the plane-wave treatment, we get $\Delta \chi^2_\mathrm{MH}$ $\approx$ 19.5, implying that we could distinguish the MH with a confidence level of nearly 4 $\sigma$, as shown in the upper panel of Fig. \ref{Dchi_SigmaE}.
However, if $\sigma_\mathrm{wp} = 0.02$, the sensitivity will be reduced due to the damping of oscillations and $\Delta \chi^2_\mathrm{MH}$ will drop to around 3.35, as shown in the bottom panel of Fig. \ref{Dchi_SigmaE}.

\begin{figure}[!htbp]
\centering{
 \includegraphics[scale=0.43]{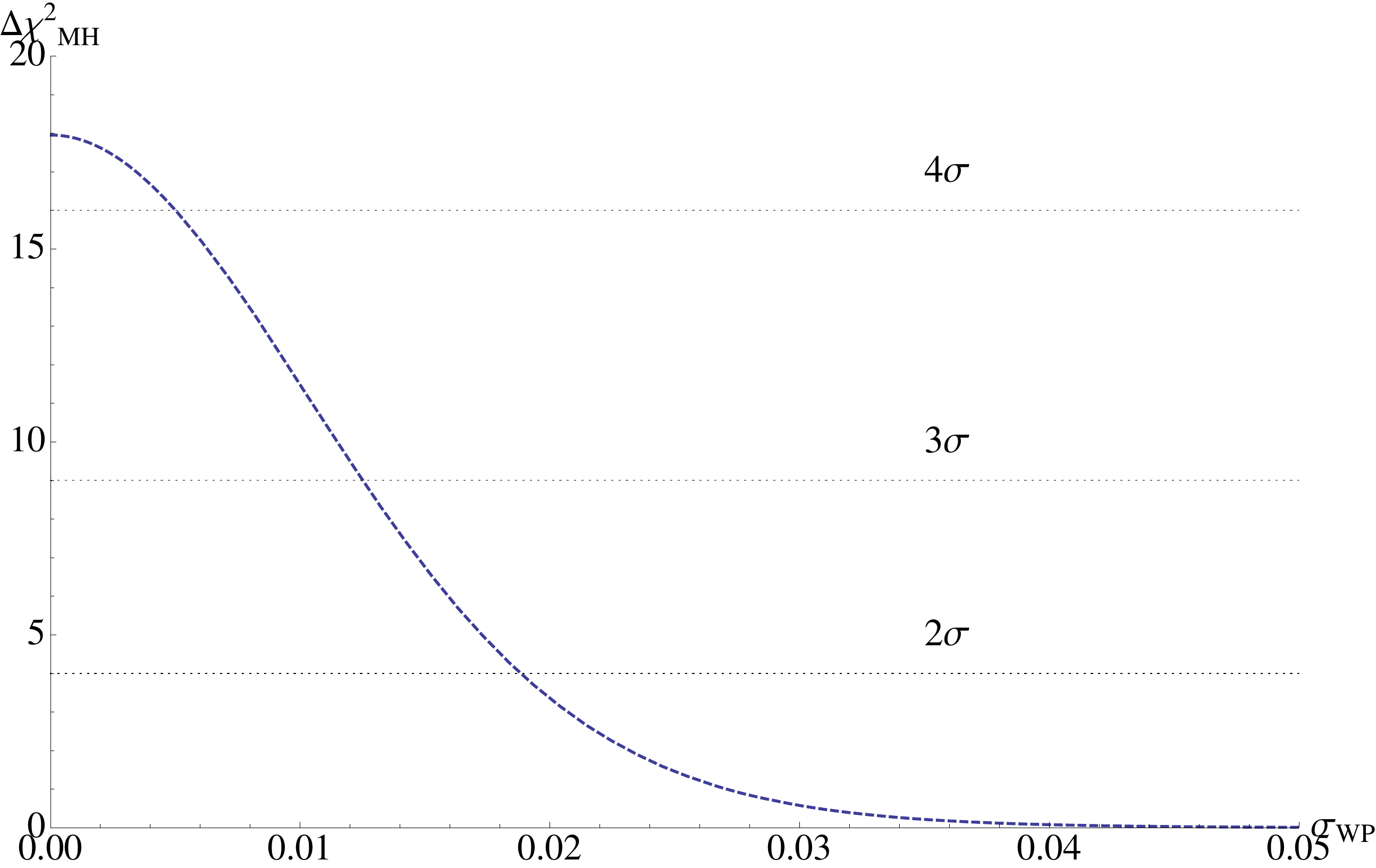} 
} 
\caption{$\Delta \chi^2_\mathrm{MH}$ vs. $\sigma_\mathrm{wp}$. The experimental setup is described in the text.}
\label{Dchi_VS_SigmaE}
\end{figure}

The solid (black) curve in Fig. \ref{Dchi_VS_SigmaE} further shows the variation of $\Delta \chi^2_\mathrm{MH}$ as a function of $\sigma_\mathrm{wp}$. It shows that $\Delta \chi^2_\mathrm{MH}$ drops rapidly with $\sigma_\mathrm{wp}$, to become smaller than 3 $\sigma$ C.L. as $\sigma_\mathrm{wp} > 0.012$. In this case, it is difficult to determine MH
.

In the rest of this section we will investigate how to increase the sensitivity to MH by improving the experimental setup.

\subsection{Consequences of energy resolution and statistics}

\footnote{The x-axes in most of the following figures just represent the ``true-value'' of $\sigma_\mathrm{wp}$. 
}References \cite{Zhan:2009rs, Petcov, Takaesu:2013wca} suggest that the sensitivity of MH measurement depends on the detector resolution. As shown in Eq. (\ref{nu_flux}), the observed $\bar{\nu_e}$ flux depends on the detector response function $R$ and energy resolution $\delta_E$, which are defined as:
\begin{align}\label{response}	
 & R(E - E_\mathrm{vis}, \delta E) = \dfrac{1}{\sqrt{2\pi} \delta E_\mathrm{vis}} \mathrm{exp}\{-\dfrac{(E-E_\mathrm{vis})^2}{2(\delta E_\mathrm{vis})^2} \}, \\
 & \mbox{where the detector energy resolution is parameterized as } \notag \\
 & \dfrac{\delta E_\mathrm{vis}}{E} = \sqrt{\left(\dfrac{a}{\sqrt{E_\mathrm{vis}/\mathrm{MeV}}}\right)^2+b^2}. \label{resolution}
\end{align}
In the previous subsection, we assumed that $a$ = 0.03 and $b$ = 0 in order to achieve a 3\% detector resolution. 

\begin{figure}[!htbp]
\centering{
 \includegraphics[scale=0.75]{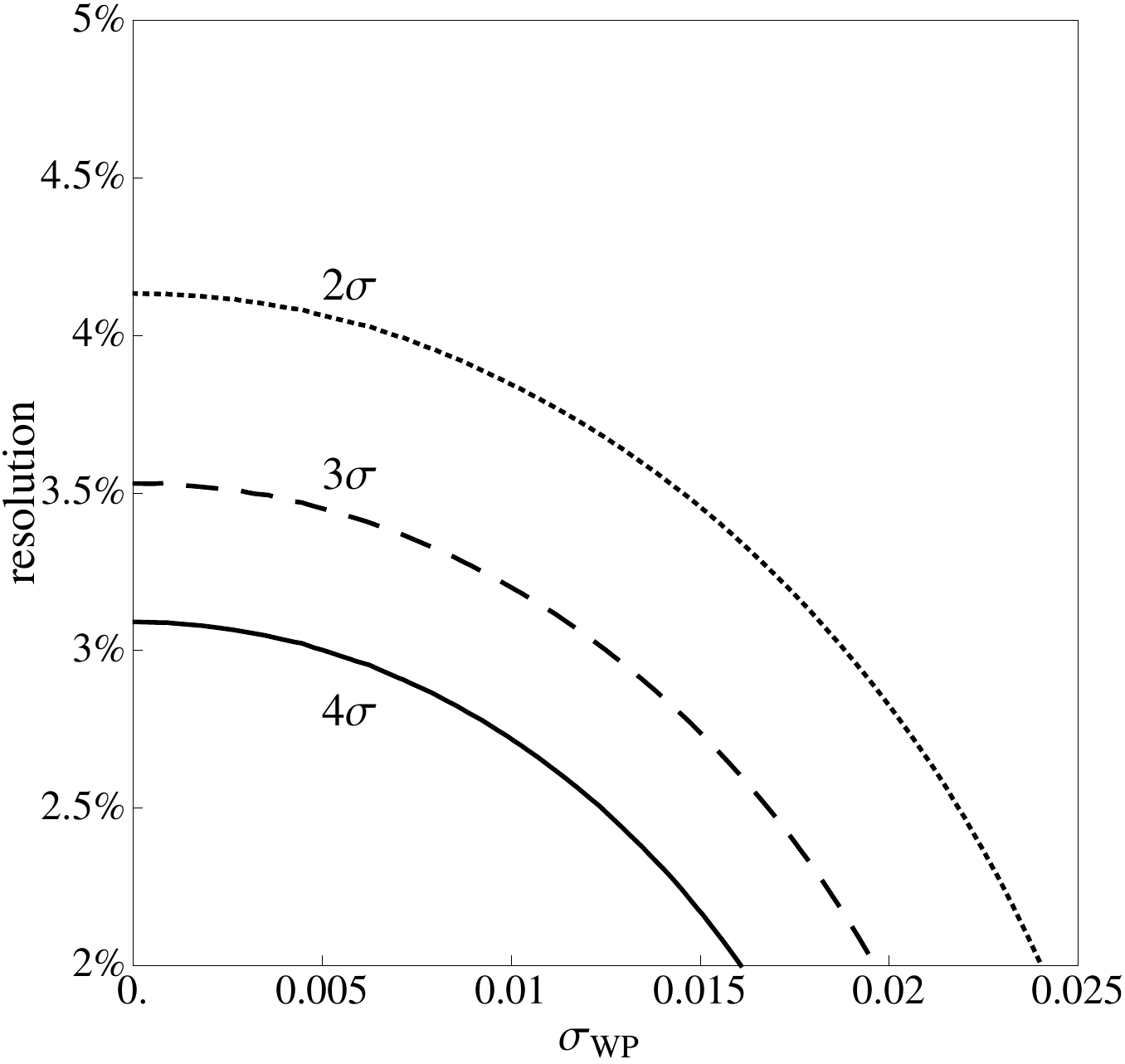} 
} 
\caption{Energy resolution required to resolve mass hierarchy at the 2 $\sigma$ (dotted), 3 $\sigma$ (dashed) and 4 $\sigma$ (solid) confidence levels, as a function of $\sigma_\mathrm{wp}$. Other details of the experimental setup are described in the text.}
\label{Res_VS_SigmaE}
\end{figure}

As mentioned in the footnote in Section 2, the detector resolution is different from the energy uncertainty of detection at the microscopic level. The detector resolution is determined by the properties of the macroscopic detector, which should be taken into account even in the plane-wave assumption. Similar to the decoherence effect due to $\sigma_\mathrm{wp}$, a poor energy resolution can also destroy the measured oscillation effect. At $L$ = 53 km, the $\theta_{13}$ oscillation could be smeared out with a finite energy resolution, and the MH information could be destroyed, particularly in the case of large $\sigma_\mathrm{wp}$. 
Fig. \ref{Res_VS_SigmaE} shows that if $\sigma_\mathrm{wp} > 0.02 $, which is allowed by the Daya Bay and KamLAND data, the detector resolution (or the parameter $a$ in Eq. \ref{resolution}), must be better than 3\% in order to determine the MH with a C.L. of more than 2 $\sigma$. 

\begin{figure}[!htbp]
\centering{
 \includegraphics[scale=0.75]{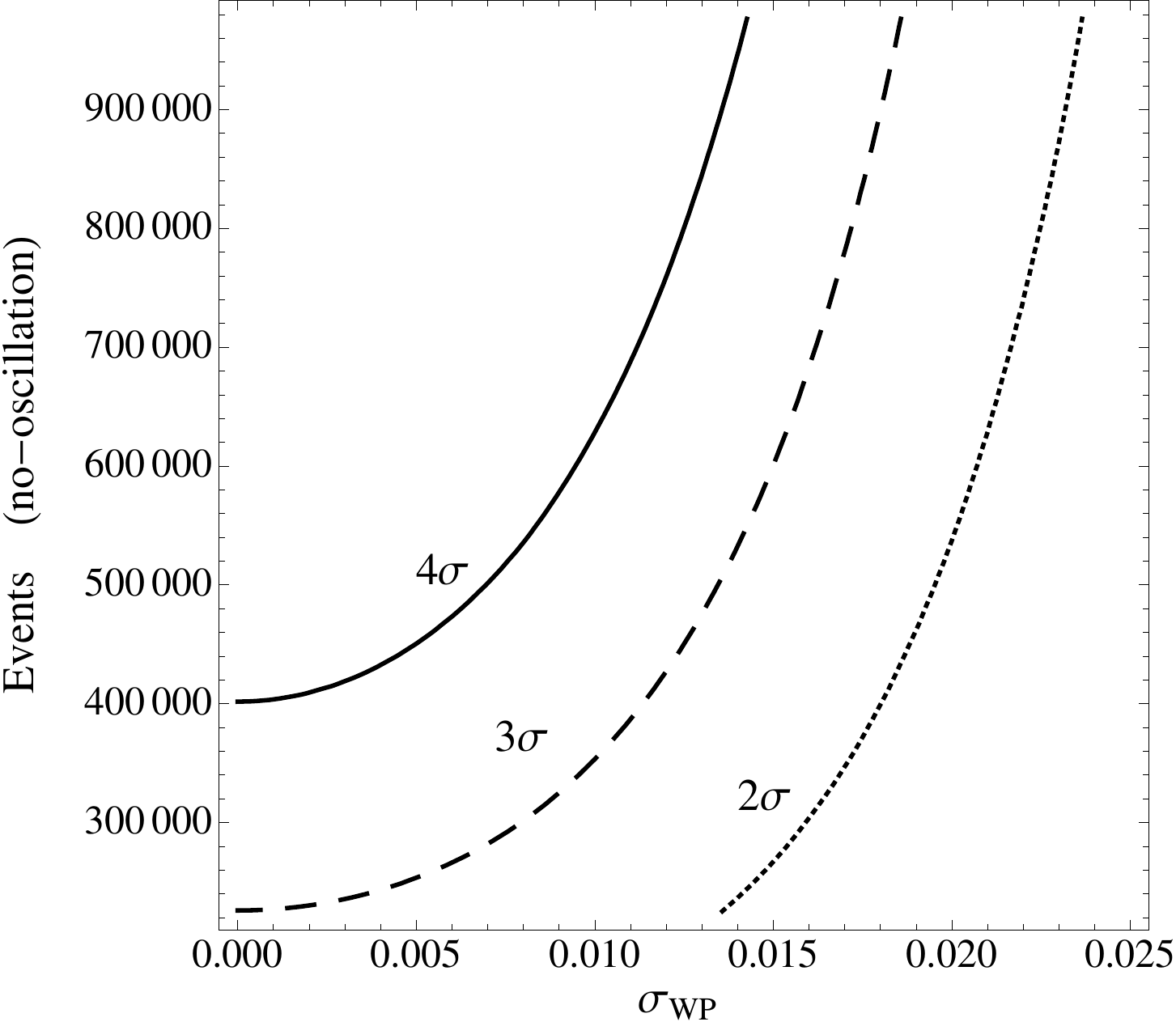} 
} 
\caption{The number of neutrino events (no-oscillation) required to resolve MH at 2 $\sigma$ (dotted), 3 $\sigma$ (dashed) and 4 $\sigma$ (solid) C.L., as a function of $\sigma_\mathrm{wp}$. Other details of the experimental setup are described in the text.}
\label{Time_VS_SigmaE}
\end{figure}

On the other hand, we can also improve the MH sensitivity by collecting more data. Fig. \ref{Time_VS_SigmaE} shows the impact of statistics on the measurement of MH as a function of $\sigma_\mathrm{wp}$, assuming a detector energy resolution of 3\%, as suggested by references \cite{Yifang, Zhan:2009rs}.  Much longer run time would be required for a 2 $\sigma$ C.L. measurement if $\sigma_\mathrm{wp}$ is larger than 0.01.

\subsection{The optimal baseline}
Without considering the wave-packet impact, 50 $\sim$ 60 km is the ideal location to measure mass hierarchy for reactor neutrino experiments \cite{Qian:2012xh, Yifang}. However, in the presence of significant wave-packet impact, longer baseline would lead to larger damping of the oscillation amplitude. Although reducing the baseline will lead to a loss of maximal phase difference between the NH and IH oscillation curves, this can save back part of the oscillation. Therefore, the optimal baseline of measuring MH could be shorter than 50 km, depending on the value of $\sigma_\mathrm{wp}$. 

\begin{figure}[!htbp]
\centering{
  \includegraphics[scale=0.47]{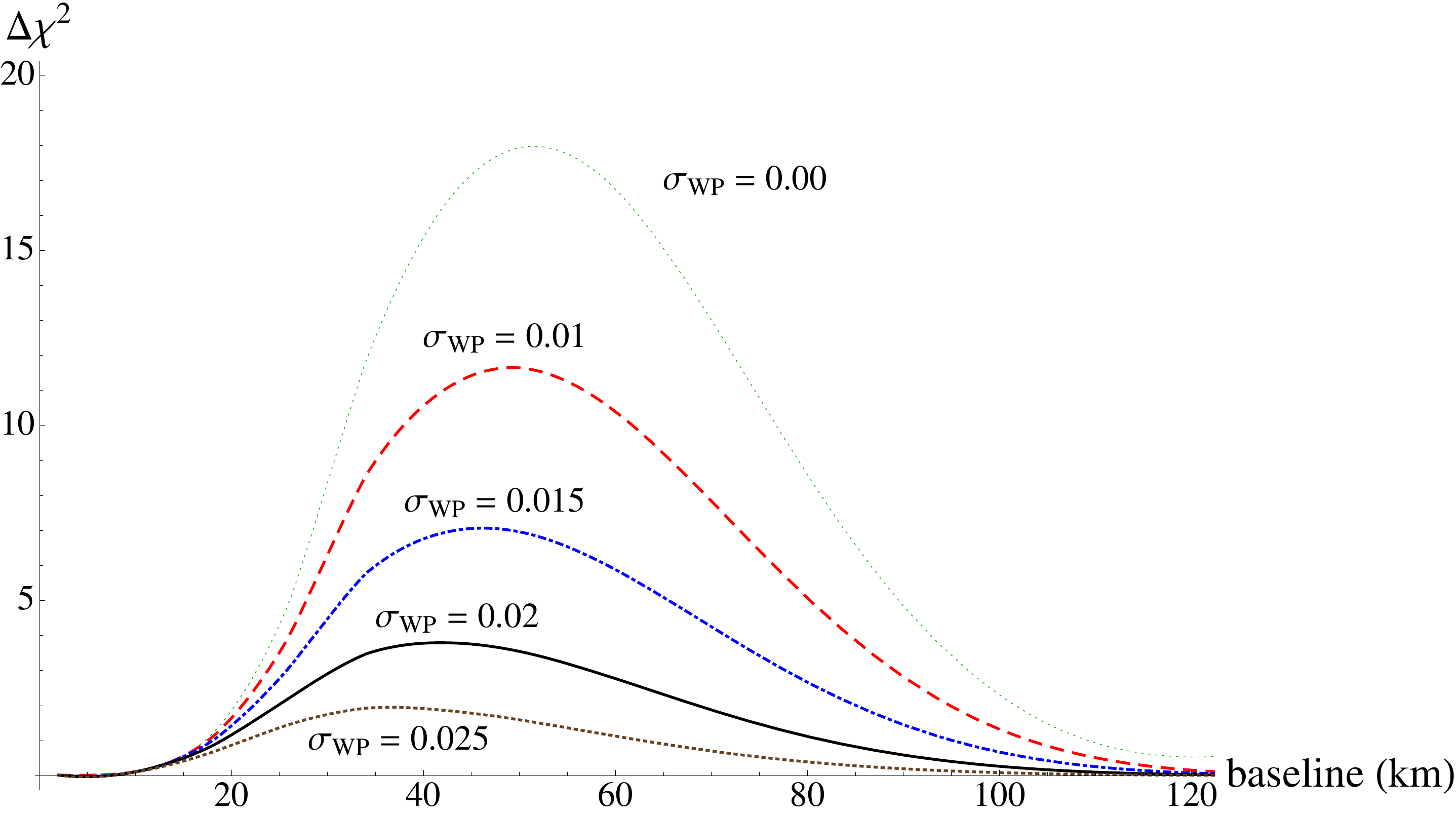}
}  
\caption{$\Delta \chi^2_\mathrm{MH}$ vs. baseline, assuming the true value of $\sigma_\mathrm{wp}$ = 0 (green dotted), 0.01 (red dashed), 0.015 (blue dot-dashed), 0.02 (black solid thick) and 0.025 (brown dotted thick) respectively. }
\label{BaselineChi2Dist}
\end{figure}

\begin{figure}[!htbp]
\centering{
 \includegraphics[scale=0.53]{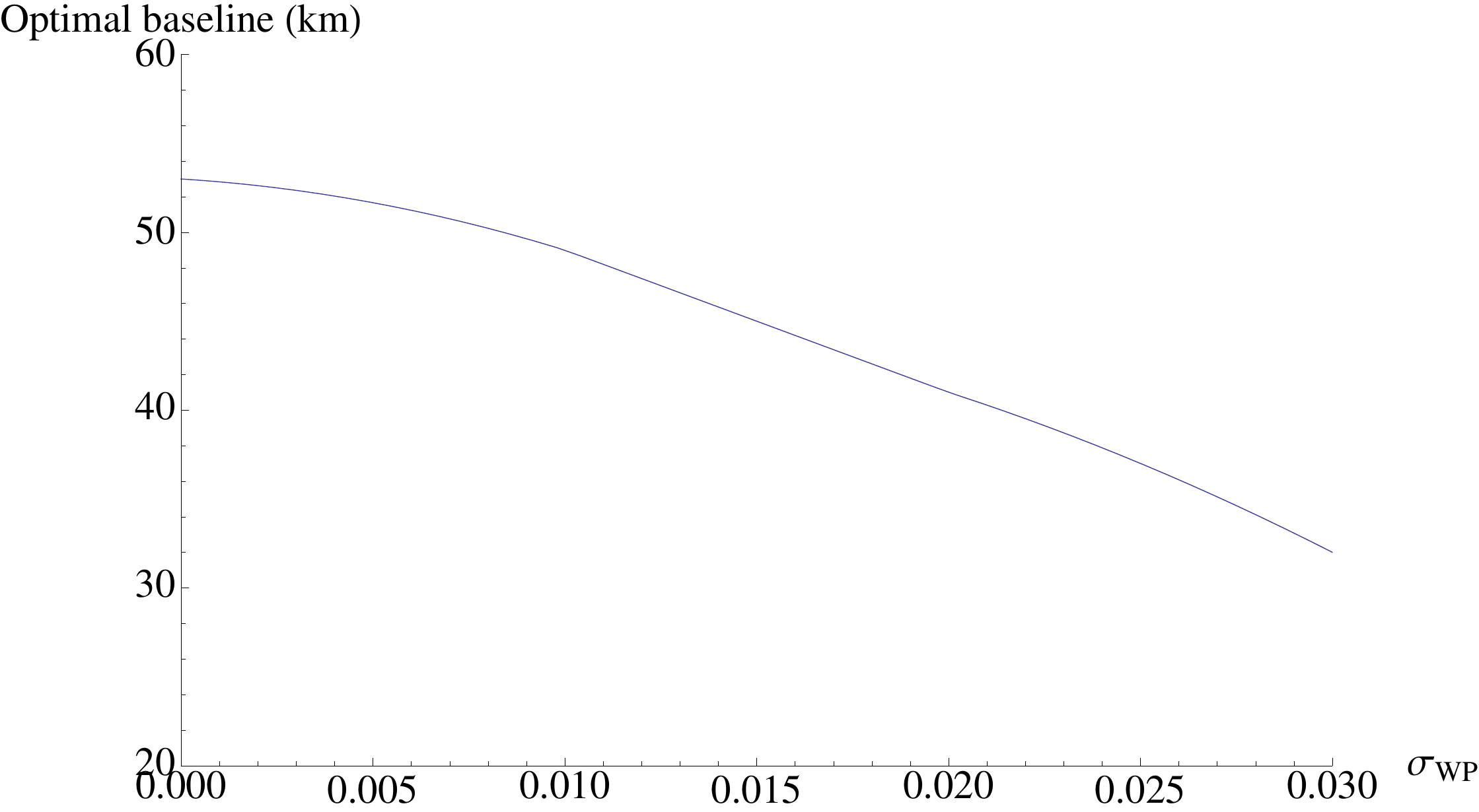} 
} 
 \caption{The optimal baseline of measuring MH as a function of $\sigma_\mathrm{wp}$.}
 \label{OptimalBaseline}
\end{figure}

Fig. \ref{BaselineChi2Dist} shows the $\Delta \chi^2_\mathrm{MH}$ as a function of baseline for different values of $\sigma_\mathrm{wp}$. In the case of the plane-wave limit ($\sigma_\mathrm{wp} = 0$), the MH can be distinguished with a confidence level of nearly 4 $\sigma$ at 53 km. However, as $\sigma_\mathrm{wp}$ increases, the maximum $\Delta \chi^2_\mathrm{MH}$ shifts to shorter baseline
. Fig. \ref{OptimalBaseline} further shows the value of optimal baseline as a function of $\sigma_\mathrm{wp}$.

\section{Conclusion}

The wave-packet impacts on current reactor neutrino oscillation and neutrino mass hierarchy measurements have been discussed. Our analyses show that the wave-packet treatment would not lead to significant modifications of the oscillation parameters from the Daya Bay and KamLAND results based on plane-wave assumptions. 
Moreover, our analyses also suggest that the energy uncertainty parameter $\sigma_\mathrm{wp}$ can be $\sim$ $O(0.1)$. 

The decoherence and dispersion effects depend not only on the initial neutrino energy uncertainty, but also the values of $\Delta m^2$ and baseline. Since the measurement of the neutrino MH in medium baseline reactor neutrino experiments relies on the fast $\Delta m^2_{32}$ oscillations, the decoherence and dispersion effects could be significant and make it more difficult. 
We found that even if $\sigma_\mathrm{wp}$ is just around 0.02, the sensitivity of MH measurement would be largely reduced. The optimal baseline shifts to smaller value as $\sigma_\mathrm{wp}$ increases,  due to the damping of oscillation amplitudes by wave-packet effects.

We have to emphasize that we are not suggesting that the wave-packet impact would be so large as to make the mass hierarchy measurement in medium baseline reactor neutrino experiments impossible. 
We point out that the plane-wave model of neutrno oscillation is only an approximation, and the wave-packet treatment is more general. While the wave-packet effect is insignificant in the current reactor neutrino oscillation experiments, its impact on future oscillation experiments needs to be determined.

\section*{Acknowledgement}
The authors thank to Dmitry Naumov, Maxim Gonchar, Emilio Ciuffoli, Jarah Evslin, Kam Biu Luk, Yu Feng Li and Alex E. Bernardini for discussions and suggestions.

This work is partially supported by grants from the Research Grant Council of the Hong Kong Special Administrative Region, China (Project Nos.  CUHK 1/07C and CUHK3/CRF/10), and the CUHK Research Committee Group Research Scheme 3110067.

\appendix
\section{Derivation of the transition probabilities}
In order to derive the neutrino oscillation probability, we substitute Eq. (\ref{E_expand}) into Eq. (\ref{wave_packet}) to obtain
\begin{align}
 |\nu_i(z,t) \rangle & = \dfrac{1}{\sqrt{1 + i \dfrac{t}{\tau_i}}} \left( \dfrac{\sigma_\nu}{\sqrt{\pi}} \right)^{1/2} \mathrm{exp} \left( i(p_\nu z-E_i(p_\nu)t) - \dfrac{\sigma_\nu^2}{2}\dfrac{(z-v_i(p_\nu)t)^2}{1 + i \dfrac{t}{\tau_i}} \right)|\nu_i \rangle. \label{wave_packet_2} \\
                     & = \dfrac{1}{\sqrt{1 + i \dfrac{t}{\tau_i}}} \left( \dfrac{\sigma_\mathrm{wp} E (p_\nu)}{\sqrt{\pi}} \right)^{1/2} \mathrm{exp} \left( i(p_\nu z-E_i(p_\nu)t) -\dfrac{(\sigma_\mathrm{wp} E(p_\nu))^2}{2}\dfrac{(z-v_i(p_\nu)t)^2}{1 + i \dfrac{t}{\tau_i}} \right)|\nu_i \rangle, \label{wave_packet_3} \\
 \mbox{where } \sigma_\mathrm{wp} & = \dfrac{\sigma_\nu}{E_i(p_\nu)} \approx \dfrac{\sigma_\nu}{E(p_\nu)}, \qquad \tau_i = \dfrac{(p_\nu^2+m_i^2)^{3/2}}{ m_i^2 \sigma_\nu^2} = \dfrac{E_i (p_\nu)^3}{ m_i^2 \sigma_\nu^2} \approx \dfrac{E (p_\nu)}{ m_i^2 \sigma_\mathrm{wp}^2}, \notag
\end{align}
which is a moving Gaussian wave packet with dispersion. Here, we have used the approximation $E_i(p_\nu) \approx p_\nu \approx E(p_\nu)$, where $E(p_\nu)$ is the neutrino energy measured by the detector. Eq. (\ref{wave_packet_3}) shows that the propagating state $|\nu_i(z,t) \rangle$ exhibits dispersion -- a spreading of the wave packet in space. Both the group velocity and the rate of dispersion depend on the neutrino mass. The heavier the mass, the smaller is the group velocity and the higher is the rate of dispersion. 

Let $|\nu_\alpha \rangle$ ($\alpha = e, \mu, \ldots$) represent the neutrino flavor states. Since the flavor eigenstates are superpositions of mass eigenstates $|\nu_i \rangle$, from Eq. (\ref{wave_packet_3}), the time evolution of a flavor state is given by
\begin{align}\label{flavor_state_trans}
 |\nu_\alpha(z,t) \rangle = & \sum_{i} U_{\alpha i}^\ast |\nu_i(z,t) \rangle \notag \\
                          = & \sum_{i}\dfrac{U_{\alpha i}^\ast}{\sqrt{1 + i \dfrac{t}{\tau_i}}} \left( \dfrac{\sigma_\mathrm{wp} E (p_\nu)}{\sqrt{\pi}} \right)^\frac{1}{2} \mathrm{exp} \left( i(p_\nu z-E_i(p_\nu)t) - \dfrac{(\sigma_\mathrm{wp} E(p_\nu))^2}{2}\dfrac{(z-v_i(p_\nu)t)^2}{1 + i \dfrac{t}{\tau_i}} \right) \notag \\
                          & \left(\sum_{\beta = e, \mu, \ldots}U_{\beta i} |\nu_\beta \rangle \right).
\end{align}
Then the transition probability of $\nu_\alpha \rightarrow \nu_\beta$ at a distance $L$ and time $t$ is given by

\begin{align}\label{Prob}
 P_{\alpha \beta}(L,t) = & |\langle \nu_\beta|\nu_\alpha(L,t) \rangle|^2 \notag\\
                                      = & \left| \sum_{i}\dfrac{U_{\alpha i}^\ast U_{\beta i}}{\sqrt{1 + i \dfrac{t}{\tau_i}}} \left( \dfrac{\sigma_\nu}{\sqrt{\pi}} \right)^\frac{1}{2} \mathrm{exp} \left( i(p_\nu L-E_i(p_\nu)t) - \sigma_\nu^2\dfrac{(L-v_i(p_\nu)t)^2}{1 + i \dfrac{t}{\tau_i}} \right) \right|^2 \notag \\
                                      = & \dfrac{\sigma_\mathrm{wp} E (p_\nu)}{\sqrt{\pi}} \sum_{i}\sum_{j} \dfrac{U_{\alpha i}^\ast U_{\beta i}U_{\alpha j}U_{\beta j}^\ast}{\sqrt{(1 + i \dfrac{t}{\tau_i})(1 - i \dfrac{t}{\tau_j})}} \mathrm{exp}\left(-i(E_i(p_\nu)-E_j(p_\nu))t\right) \cdot \notag \\
                                        & \mathrm{exp} \left( - \sigma_\mathrm{wp}^2 (E (p_\nu))^2 \left(\dfrac{(L-v_i(p_\nu)t)^2}{1 + i \dfrac{t}{\tau_i}}+\dfrac{(L-v_j(p_\nu)t)^2}{1 - i \dfrac{t}{\tau_j}} \right)\right),
\end{align}
which is a function of both time ($t$) and distance ($L$).

In an oscillation experiment, the neutrino is detected at a fixed baseline $L$ but the time is not measured. In order to obtain the oscillation probability as a function of the baseline, the time has to be integrated out in Eq. (\ref{Prob}). Since reactor neutrinos propagate almost at the speed of light, $P_{\alpha \beta}(L,t)$ is non-zero only around $t \sim L$. The transition probability is non-zero only within a small time window $\Delta t$ around a time $t_L$ where 
\begin{equation}\label{t_L}
L=\frac{v_i+v_j}{2}t_L, \qquad \rightarrow t_L=\frac{2L}{v_i+v_j} \sim L.
\end{equation}
On the other hand, the size of $\Delta t$ is constrained by the spatial width of the wave packet which is typically much smaller than the baseline $L$, which means that $\Delta t$ $\ll$ $t_L$. Moreover,
\begin{align}
 \dfrac{\partial \left(\dfrac{t}{\tau_i}\right)}{\partial t} = \dfrac{1}{\tau_i} & \approx \dfrac{m_i^2}{E(p_\nu)} \cdot \sigma_\mathrm{wp}^2 \sim 0, \label{approx_der1} \\
 \left|\dfrac{\partial (z-v_i(p_\nu)t)}{\partial t}\right| & = |v_i(p_\nu)|. \label{approx_der3}
\end{align}
Eq. (\ref{approx_der1}) shows that the factor ($t/\tau_i$) changes slowly with the variable $t$. Within a small $\Delta t$, these terms can be treated as constants. Therefore, within the small integration region which is constrained by the width of the wave packet, it is acceptable to approximate $t = t_L \sim z$ for this factor,  
\begin{align}
 \dfrac{t}{\tau_i} = \dfrac{m_i^2 \sigma_\mathrm{wp}^2t}{E(p_\nu)} & \sim \dfrac{m_i^2 \sigma_\mathrm{wp}^2}{E(p_\nu)}z. \label{approx1} 
\end{align}
However, since Eq. (\ref{approx_der3}) shows that the derivative of $(z-v_i(p_\nu)t)$ is not negligible even in a small region. We did not use the same approximation in the factors $(z-v_i(p_\nu)t)$ and $(z-v_j(p_\nu)t)$. 
Therefore, the integral of Eq. (\ref{Prob}) can be approximated as,
\begin{align}\label{Appendix_Prob_inf}
 P_{\alpha \beta}(z) & = \int P_{\alpha \beta}(z,t) dt \notag \\
                                        & \approx\int \dfrac{\sigma_\mathrm{wp} E(p_\nu)}{\sqrt{\pi}} \sum_{i,j} \dfrac{U_{\alpha i}^\ast U_{\beta i}U_{\alpha j}U_{\beta j}^\ast e^{(-i(E_i(p_\nu)-E_j(p_\nu)) z)}}{\sqrt{(1 + i \dfrac{m_i^2 \sigma_\mathrm{wp}^2z}{E(p_\nu)})(1 - i \dfrac{m_j^2 \sigma_\mathrm{wp}^2z}{E(p_\nu)})}} \notag \\
                                        & \mathrm{exp} \left( - \dfrac{\sigma_\mathrm{wp}^2 (E(p_\nu))^2}{2} \left(\dfrac{(z-v_i(p_\nu)t)^2}{1 + i \dfrac{m_i^2 \sigma_\mathrm{wp}^2z}{E(p_\nu)}}+\dfrac{(z-v_j(p_\nu)t)^2}{1 - i \dfrac{m_j^2 \sigma_\mathrm{wp}^2z}{E(p_\nu)}} \right)\right)dt \notag \\
                                        & \approx \sum_{i,j} (U_{\alpha i}^\ast U_{\beta i}U_{\alpha j}U_{\beta j}^\ast \mathrm{exp}(\dfrac{-i\Delta m_{ij}^2 z}{2E(p_\nu)}))
                                         \cdot \left[\dfrac{1}{\sqrt{1 + i \dfrac{\Delta m_{ij}^2 \sigma_\mathrm{wp}^2z}                                        {2E(p_\nu)}}}\mathrm{exp}\left(- \dfrac{\dfrac{(\Delta m_{ij}^2\sigma_\mathrm{wp} z)^2}{16 (E(p_\nu))^2}}{1+i\dfrac{\Delta m_{ij}^2 \sigma_\mathrm{wp}^2 z}{2E(p_\nu)}} \right) \right] \notag \\
                                        & \cdot \mathrm{exp}\left[ -\dfrac{(\dfrac{(\Delta m_{ij}^2)^2}{E(p_\nu)})^4}{8\sigma_\mathrm{wp}^2} + \dfrac{(\dfrac{(\Delta m_{ij}^2)^2}{E(p_\nu)})^4}{16(1 + i \dfrac{\Delta m_{ij}^2 \sigma_\mathrm{wp}^2z}{2E(p_\nu)})\sigma_\mathrm{wp}^2} \right].
\end{align}
In the last step of Eq. (\ref{Appendix_Prob_inf}), we have ignored the terms proportional to $\left( \dfrac{m_i^2 m_j^2}{E(p_\nu)^2}\right)$, since they are expected to be negligible for ultrarelativistic neutrinos. In the case of reactor neutrino experiments, the $\bar{\nu_e}$ survival probability is given by
 \begin{align}\label{Appendix_Pee}
    P_{\bar{e}\bar{e}} = & 1 - \frac{1}{2}\mathrm{cos}^4(\theta_{13})\mathrm{sin}^2(2\theta_{12})[1 - (\dfrac{1}{1+y_{21}^2})^{\frac{1}{4}}\mathrm{exp}(-\lambda_{21})\mathrm{exp}(-\gamma_{21})\mathrm{cos}(\phi_{21})] \notag \\
          & - \frac{1}{2}\mathrm{sin}^2(2\theta_{13})\mathrm{cos}^2(\theta_{12})[1 - (\dfrac{1}{1+y_{31}^2})^{\frac{1}{4}}\mathrm{exp}(-\lambda_{31})\mathrm{exp}(-\gamma_{31})\mathrm{cos}(\phi_{31})] \notag \\
          & - \frac{1}{2}\mathrm{sin}^2(2\theta_{13})\mathrm{sin}^2(\theta_{12})[1 - (\dfrac{1}{1+y_{32}^2})^{\frac{1}{4}}\mathrm{exp}(-\lambda_{32})\mathrm{exp}(-\gamma_{32})\mathrm{cos}(\phi_{32})], \\
\mbox{where }& \sigma_\mathrm{wp} \equiv \dfrac{\sigma_\nu}{E}, \qquad y_{ij} = \dfrac{\Delta m^2_{ij}L}{2E} \cdot \sigma_\mathrm{wp}^2, \notag \\
             & \lambda_{ij} = \dfrac{\dfrac{1}{16}\dfrac{((\Delta m_{ij}^2)^2 \sigma_\mathrm{wp}^2 L^2)}{E^2}}{1+y_{ij}^2}, \qquad \gamma_{ij} = \dfrac{\dfrac{1}{16}\dfrac{(\Delta m_{ij}^2)^2}{E^4}}{1+y_{ij}^2}\cdot\dfrac{1}{\sigma_\mathrm{wp}^2} \notag \\ 
             & \phi_{ij} = \dfrac{\Delta m^2_{ij} L}{2E} + \left(\frac{1}{2} \mbox{tan}^{-1} (y_{ij}) - \dfrac{\dfrac{1}{16}\dfrac{((\Delta m_{ij}^2)^2 \sigma_\mathrm{wp}^2 L^2)}{E^2} \cdot y_{ij}}{1+y_{ij}^2}\right). \notag
\end{align}
The damping factor $\mathrm{exp}(-\gamma_{ij})$ corresponds to an decoherence effect from delocalization, which is neglected in the main article. This is because in most circumstances, $\gamma_{ij} \approx 0$. The details of $\gamma_{ij}$ and the delocalization decoherence effect will be discussed in Appendix B.

\section{The decoherence effect from delocalization}
The decoherence effect discussed in Section 2 is due to the separation of different neutrino wave packets. With larger values of $\sigma_\mathrm{wp}$, the decoherence effect would be more significant. On the other hand, there exists another decoherence effect which is due to the delocalization of the production and detection processes. Different from what we have studied above, the decoherence effect from delocalization will become significant only if $\sigma_\mathrm{wp}$ is extremely small, or the spatial uncertainty $\sigma_x$ is large. In fact, in neutrino oscillations, one of the coherence conditions is that the intrinsic production (and also detection) energy uncertainties are much larger than the energy differences between different mass eigenstates $\Delta E_{ij} = E_i -E_j$ ($E_i$ is the energy of mass eigenstate $|\nu_i \rangle$) \cite{Akhmedov:2012uu}, namely,
\begin{align}\label{condition}
 \Delta E_{ij} \equiv E_i - E_j \sim \dfrac{\Delta m_{ij}^2}{E_\nu} & \ll \sigma_\nu \equiv E_\nu\sigma_\mathrm{wp}, \notag\\
 \mbox{    which means } \quad \sigma_x & \ll L^\mathrm{osc}_{ij} \quad (\sigma_\nu \sim 1/\sigma_x).
\end{align}
Eq. (\ref{condition}) implies that in order to measure the interferences between different mass eigenstates, the spatial uncertainty $\sigma_x$, has to be much smaller than the oscillation length $L^\mathrm{osc}_{ij}$. 

Eq. (\ref{Prob_inf1}) is just an approximate neutrino oscillation probability formula and it does not describe the decoherence effect from delocalization. More precisely, the flavor transition probability in Eq. (\ref{Prob_inf1}) should be multiplied by an additional factor
\begin{align}\label{Prob_complete}
 & \mathrm{exp}(-\gamma_{ij}), \\
 \mbox{where }\qquad \gamma_{ij} & = \dfrac{\pi^2}{(L^\mathrm{osc}_{ij})^2 E^2 \sigma_\mathrm{wp}^2} = \dfrac{\pi^2}{(1+y_{ij}^2)}\cdot\dfrac{\sigma_x^2}{(L^\mathrm{osc}_{ij})^2}, \notag
\end{align}
With these delocalization terms also taken into account, a more complete $\bar{\nu_e}$ survival probability should be written as Eq. (\ref{Appendix_Pee}) in the previous section. The damping factor in Eq. (\ref{Prob_complete}) corresponds to the decoherence effect from delocalization. If $\gamma_{ij}$ $\sim$ 0, the modifications from delocalization are negligible. In this case Eq. (\ref{Appendix_Pee}) just reduces to Eq. (\ref{Prob_inf1}). $\gamma_{ij}$ is proportional to $\dfrac{\sigma_x^2}{(L^\mathrm{osc}_{ij})^2}$, and so the decoherence effect from delocalization matters only when $\sigma_x$ $\sim$ $L^\mathrm{osc}_{ij}$, about 1 km in the measurement of $\theta_{13}$ oscillation. In this case $\sigma_\mathrm{wp}$ is extremely small.

\begin{figure}[!htbp]
\centering
 \includegraphics[scale=0.5]{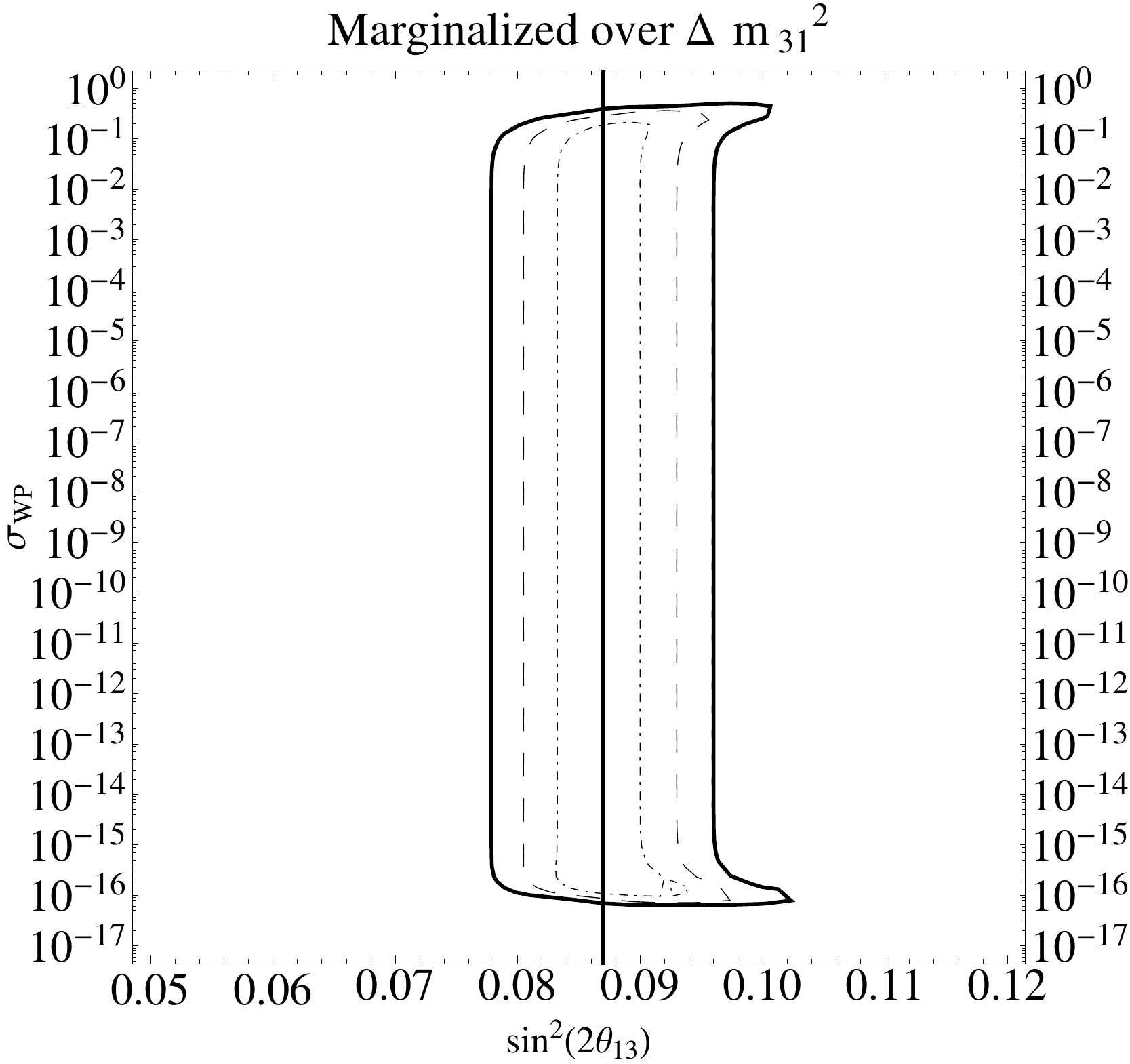}
\caption{Same as Fig. \ref{DB_t13_sigmaE}, but delocalization is taken into account and the y-axis is in log scale.}
\label{DB_t13_sigmaE_lower}
\end{figure}

If we use Eq. (\ref{Appendix_Pee}) rather than Eq. (\ref{Prob_inf1}) to do analysis, we will find that the $\gamma_{ij}$ term will offer a lower bound on $\sigma_\mathrm{wp}$ and the delocalization effect is significant only when $\sigma_\mathrm{wp}$ is extremely small, which implies a large $\sigma_x$. Fig. \ref{DB_t13_sigmaE_lower} shows our analysis of the Daya Bay data, which is similar to Fig. \ref{DB_t13_sigmaE}, but this time the delocalization term is considered.

From Fig. \ref{DB_t13_sigmaE_lower} we can see that in a large range of the parameter space, there are no modifications on the true value of sin$^2 2\theta_{13}$. This means that in most of the parameter region, wave-packet impact can be safely neglected. Moreover, Fig. \ref{DB_t13_sigmaE_lower} also suggests that only if $\sigma_\mathrm{wp}$ $\sim O(10^{-16})$, which means that $\sigma_x$ $\sim$ $O$(1 km), the decoherence effect from delocalization is significant. 

At this point, we can discuss the wave-packet impact in two different regimes. If $\sigma_\mathrm{wp}$ is large ($\sim O(10^{-1})$), since  $\lambda_{ij}$ $\propto$ $\sigma_\mathrm{wp}^2$, the damping factors exp$(-\lambda_{ij})$ in Eq. (\ref{Prob_inf1}) become significant and the decoherence effect from the separation of wave packets cannot be neglected. On the other hand, if $\sigma_\mathrm{wp}$ $\sim O(10^{-16})$ or even smaller, then the additional damping factor in Eq. (\ref{Prob_complete}) starts to dominate since $\gamma_{ij}$ $\propto$ $\sigma_\mathrm{wp}^{-2}$. In this case the decoherence effect from delocalization becomes important. Nevertheless, in most reactor neutrino experiments, the dimensions of the reactor cores and detectors are just around a few meters. It is unlikely that the spatial width of the initial neutrino wave packet would be larger than 1 km.

However, in most reactor experiments, including current ones such as Daya Bay and KamLAND, and also the proposed measurements of neutrino mass hierarchy at medium baseline, the delocalization terms $\mathrm{exp}(-\gamma_{ij})$ can be neglected. 
This is because $\dfrac{\pi^2}{(L^\mathrm{osc}_{ij})^2 E^2 \sigma_\mathrm{wp}^2} \ll 1$ (or, the spatial uncertainty $\sigma_x \ll L^\mathrm{osc}_{ij}$).
In this case, Eq. (\ref{Prob_complete}) just reduces to Eq. (\ref{Prob_inf1}). Therefore, we neglected the decoherence effect from delocalization in our study and performed the simulations and analyses with Eq. (\ref{Prob_inf1}). \\

\bibliographystyle{elsarticle-num}

\bibliography{reference_dec.bib} 

\end{document}